\documentclass[twocolumn,superscriptaddress]{revtex4-1}
\usepackage{graphicx}
\usepackage{units}
\usepackage{multirow}
\usepackage{bigstrut}
\usepackage{overpic}
\usepackage{array}
\usepackage{color}
\usepackage{amsmath}
\usepackage{xspace}
\usepackage{amsfonts}
\usepackage{subfigure}
\usepackage{verbatim}
\usepackage{footmisc}
\usepackage{color}
\usepackage{epstopdf}
\usepackage{appendix}
\RequirePackage{lineno}
\usepackage[colorlinks,linkcolor=blue,anchorcolor=blue,citecolor=blue]{hyperref}
\newcommand{\zcsm}{Z_{cs}(3985)^-}

\newcommand{\dsm}{D_s^-}

\newcommand{\dsstm}{D_s^{*-}}
\newcommand{\DDsone}{D_s^-D^{*0}}
\newcommand{\DDstwo}{D_s^{*-} D^{0}}
\newcommand{\dstzero}{D^{*0}}

\newcommand{\dzero}{D^{0}}

\newcommand{\ee}{e^+e^-}
\newcommand{\kaonp}{K^+}
\newcommand{\kaonm}{K^-}
\newcommand{\kshort}{K_S^0}

\newcommand{\pim}{\pi^-}
\newcommand{\mev}{\,\unit{MeV}}

\newcommand{\mevcc}{\,\unit{MeV}/c^2}
\newcommand{\gev}{\,\unit{GeV}}

\newcommand{\gevcc}{\,\unit{GeV}/c^2}
\newcommand{\invpb}{\,\unit{pb}^{-1}}
\newcommand{\br}[1]{\mathcal{B}(#1)}

\begin{document}


\title{\boldmath Observation of a Near-Threshold Structure in the $K^+$ Recoil-Mass Spectra in $e^+e^-\to K^+(D_s^- D^{*0}+D^{*-}_s D^0)$}

\date{\it \small \bf \today}

\author{
\begin{small}
M.~Ablikim$^{1}$, M.~N.~Achasov$^{10,c}$, P.~Adlarson$^{67}$, S. ~Ahmed$^{15}$, M.~Albrecht$^{4}$, R.~Aliberti$^{28}$, A.~Amoroso$^{66A,66C}$, Q.~An$^{63,50}$, ~Anita$^{21}$, X.~H.~Bai$^{57}$
% ...
\\
\vspace{0.2cm}
(BESIII Collaboration)\\
\vspace{0.2cm} {\it
$^{1}$ Institute of High Energy Physics, Beijing 100049, People's Republic of China\\
% ...
$^{4}$ Bochum Ruhr-University, D-44780 Bochum, Germany\\
% ...
$^{10}$ G.I. Budker Institute of Nuclear Physics SB RAS (BINP), Novosibirsk 630090, Russia\\
% ...
$^{15}$ Helmholtz Institute Mainz, Johann-Joachim-Becher-Weg 45, D-55099 Mainz, Germany\\
% ...
$^{21}$ Indian Institute of Technology Madras, Chennai 600036, India\\
% ...
$^{28}$ Johannes Gutenberg University of Mainz, Johann-Joachim-Becher-Weg 45, D-55099 Mainz, Germany\\
% ...
$^{50}$ State Key Laboratory of Particle Detection and Electronics, Beijing 100049, Hefei 230026, People's Republic of China\\
% ...
$^{57}$ University of Jinan, Jinan 250022, People's Republic of China\\
% ...
$^{63}$ University of Science and Technology of China, Hefei 230026, People's Republic of China\\
% ...
$^{66}$ University of Turin and INFN, INFN, I-10125, Turin, Italy\\
$^{67}$ Uppsala University, Box 516, SE-75120 Uppsala, Sweden\\
% ...
$^{c}$ Also at the Novosibirsk State University, Novosibirsk, 630090, Russia\\
% ...
}
\vspace{0.4cm}
\end{small}
}


\vspace{4cm}

\begin{abstract}
We report a study of the processes of $e^+e^-\to K^+ D_s^- D^{*0}$ and $K^+ D^{*-}_s D^0$ based on $e^+e^-$ annihilation samples
collected with the BESIII detector operating at BEPCII at five center-of-mass energies ranging from 4.628 to 4.698 GeV with a total integrated luminosity of 3.7 fb$^{-1}$.
An excess of events over the known contributions of the conventional charmed mesons is observed near the $D_s^- D^{*0}$ and $D^{*-}_s D^0$ mass thresholds in the $K^{+}$ recoil-mass spectrum for events collected at $\sqrt{s}=4.681$ GeV.
The structure matches a mass-dependent-width Breit-Wigner line shape, whose pole mass and width are determined as  $(3982.5^{+1.8}_{-2.6}\pm2.1)$ MeV/$c^2$ and $(12.8^{+5.3}_{-4.4}\pm3.0)$ MeV, respectively.
The first uncertainties are statistical and the second are systematic.
The significance of the resonance hypothesis is estimated to be 5.3 $\sigma$ over the contributions only from the conventional charmed mesons.
This is the first candidate for a charged hidden-charm tetraquark with strangeness, decaying into $D_s^- D^{*0}$ and $D^{*-}_s D^0$.
However, the properties of the excess need further exploration with more statistics.
\end{abstract}


\maketitle


Recent observations of nonstrange hidden-charm tetraquark candidates with quark content $c\bar{c}q\bar{q}'$ ($q^{(\prime)}=u$ or $d$), referred to as the $Z_c$ states, have opened a new chapter in hadron spectroscopy~\cite{XYZ_review1,XYZ_review2,XYZ_review3,XYZ_review4,XYZ_review5,XYZ_review6}.
In electron-positron annihilation, in particular, both the charged and neutral $Z_c(3900)$ and $Z_c(4020)$ have been observed at the BESIII, Belle, and CLEO experiments in a variety of decay modes~\cite{Ablikim:2013mio1,Liu:2013dau,Xiao:2013iha,Ablikim:2013wzq1,Ablikim:2013emm1,Ablikim:2013xfr1,Ablikim:2014dxl,Ablikim:2015tbp,Ablikim:2015vvn,Ablikim:2015gda}.
Assuming SU(3) flavor symmetry, one would expect the existence of strange partners to the $Z_c$, denoted as $Z_{cs}$, with quark content $c\bar{c}s\bar{q}$~\cite{Voloshin:2019ilw}.
No experimental searches for $Z_{cs}$ states have yet been reported.

The existence of a $Z_{cs}$ state with a mass lying around the $\DDsone$ and $\DDstwo$ thresholds has been predicted in several theoretical models, including tetraquark scenarios~\cite{Ebert:2008kb, Ferretti:2020ewe}, the $D_s\bar{D}^*$ molecular model~\cite{Lee:2008uy,Dias:2013qga}, the hadro-quarkonium model~\cite{Ferretti:2020ewe}, and in the initial-single-chiral-particle-emission mechanism~\cite{Chen:2013wca}.
Like the $Z_c$ states, the decay rate of the $Z_{cs}$ to open-charm final states is expected to be larger than the decay rate to charmonium final states~\cite{XYZ_review5}.
Hence, one promising method to search for the $Z_{cs}$ state is through its decays to $\DDsone$ and $\DDstwo$.

In this Letter, we report on a study of the process $e^+e^-\to K^+ D_s^- D^{*0}$ and $K^+ D^{*-}_s D^0$ [$\ee\to\kaonp(\DDsone + \DDstwo)$ for short] at center-of-mass energies $\sqrt{s}=\;$4.628, 4.641, 4.661, 4.681, and 4.698\gev. The data samples have a total integrated luminosity of $3.7~{\rm fb}^{-1}$ and were accumulated by the BESIII detector at the BEPCII collider. Details about BEPCII and BESIII can be found in Refs.~\cite{Yu:IPAC2016-TUYA01, Ablikim:2009aa, Ablikim:2019hff}.
To improve the signal-selection efficiency, a partial-reconstruction technique is implemented in which only the charged $\kaonp$ (the \emph{bachelor} $\kaonp$) and the $D_s^-$ are reconstructed.
Here and elsewhere, charge-conjugate modes are always implied, unless explicitly stated otherwise.
To improve the signal purity, we only reconstruct the decays $\dsm\to K^+K^-\pi^-$ and $K_S^0K^-$, which have large branching fractions~(BFs).
By reconstructing the $\dsm$ meson, the flavors of the missing $\dzero$ and the bachelor $\kaonp$ are fixed.
We observe an enhancement near the $\dsm\dstzero$ and $\dsstm\dzero$ mass thresholds in the $K^{+}$ recoil-mass spectrum for events collected at $\sqrt{s}=4.681\gev$ and carry out a fit to the enhancement with a possible new $Z_{cs}$ candidate, denoted as $\zcsm$, in the $\kaonp$ recoil-mass spectra at different energy points.


Monte Carlo~(MC) simulation samples are produced under a {\sc geant4}-based~\cite{geant4} framework, as detailed in Ref.~\cite{Ablikim:2018qjv}.
For the three-body nonresonant~(NR) signal process, $\ee\to\kaonp(\dsm\dstzero + \dsstm\dzero)$, the final-state particles are simulated assuming nonresonant production~\cite{Ablikim:2018qjv}.
For the simulation of the $\zcsm$ signal process, $\ee\to\kaonp\zcsm$, we let the $\zcsm$ decay into the $\dsm\dstzero$ and $\dsstm\dzero$ final states with equal rates. The $\zcsm$ state is assigned a spin parity of $1^{+}$, as the corresponding production and subsequent decay processes are both in the most favored $S$ wave.
However, other spin-parity assignments are allowed, and these are tested as systematic variations.


To identify the processes $\ee\to K^+ (\DDsone + \DDstwo)$, we reconstruct combinations of the bachelor $\kaonp$ and the decays $\dsm\to K^+K^-\pim$ or $K^0_S K^-$.
Data taken at all five center-of-mass energy points are analyzed using the same procedure, but two-thirds of the data set at $\sqrt{s}=\;$4.681~GeV was kept blinded until after the analysis strategy was established and validated~\cite{supple}.
We select events with at least four charged tracks and reconstruct the final states of $K^\pm$, $\pi^\pm$ and $K_S^0\to\pi^+\pi^-$ following the criteria in  Ref.~\cite{Ablikim:2018jun}.
For the candidate of $K_S^0$, we require its invariant mass within $0.485<M(\pi^{+}\pi^{-})<0.511\gevcc$.
For the decay $\dsm \to \kaonp\kaonm\pim$, to improve the signal purity, we only retain the $\dsm$ candidates within the Dalitz plot regions consistent with $D_s^-\to \phi\pi^-$ or $D_s^{-}\to K^{*}(892)^0K^-$ decays by requiring that the invariant masses satisfy either $M(\kaonp\kaonm)<1.05\gevcc$ or $0.850<M(\kaonp\pim)<0.930\gevcc$.

\begin{figure}[tp!]
\centering
\begin{overpic}[width=0.48\linewidth]{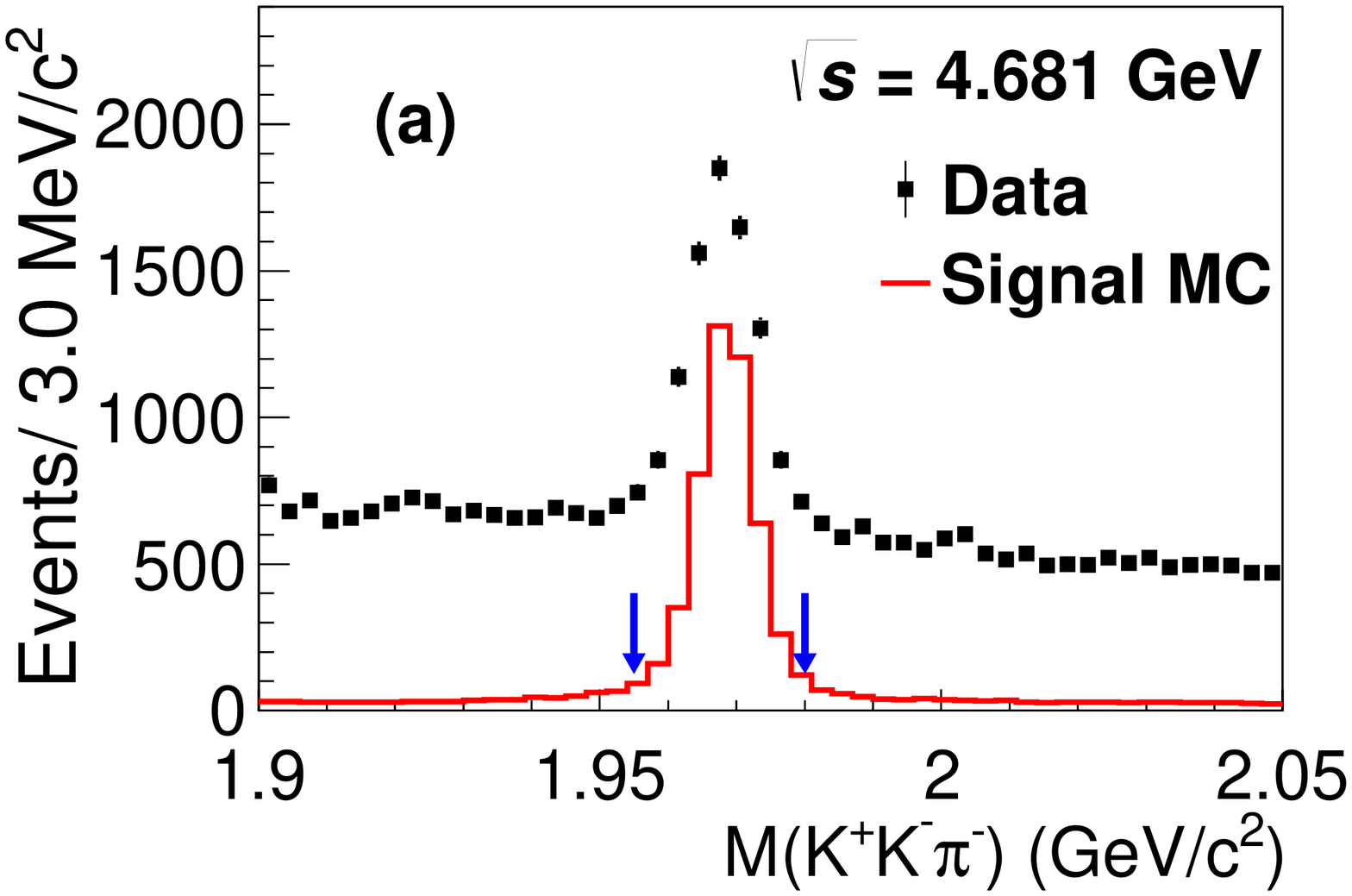}
\end{overpic}
\begin{overpic}[width=0.48\linewidth]{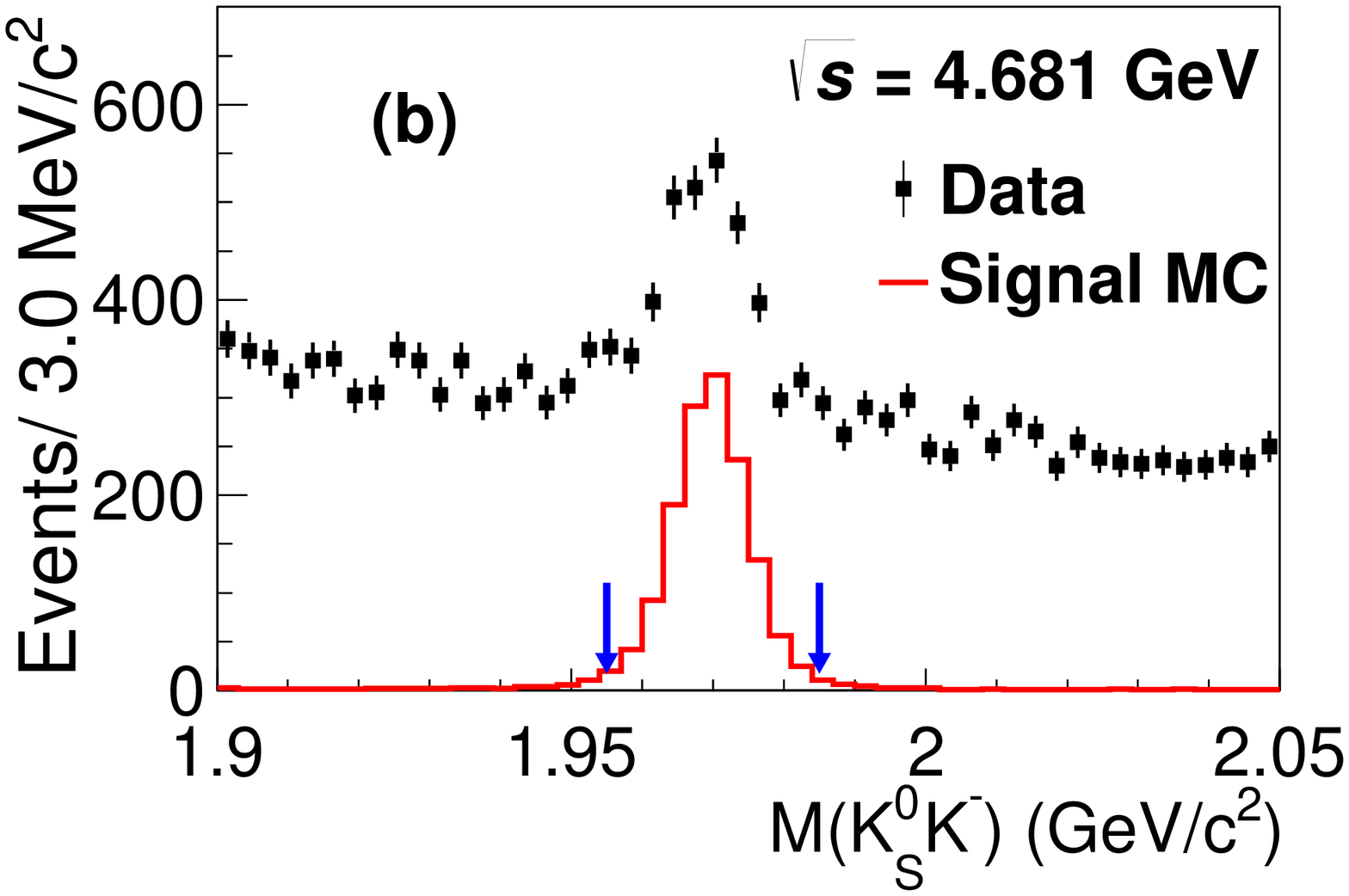}
\end{overpic}
\caption{Distributions of the invariant mass $M(K^+K^-\pim)$ (a) and $M(\kshort K^-)$ (b) in data and MC simulations at $\sqrt{s}=4.681\gev$. The $Z_{cs}(3985)^{-}$ signal MC component is normalized to the observed $\dsm$ yield in data. Arrows indicate the mass region requirements.}
\label{fig:dsmass}
\end{figure}

\begin{figure}[tp!]
\centering
\begin{overpic}[width=0.75\linewidth]{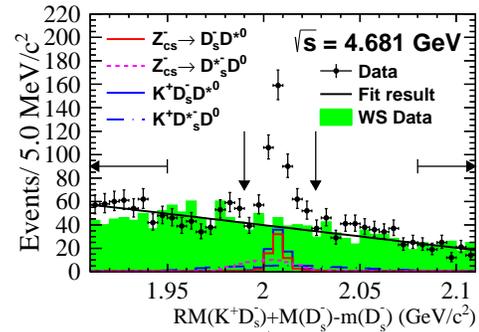}
\end{overpic}
\vspace{-0.5cm}
\caption{Distribution of the $\kaonp\dsm$ recoil mass in data and signal MC samples at $\sqrt{s}=4.681\gev$.
Horizontal arrows indicate the sidebands and vertical arrows indicate the signal region.
The magnitudes of the three-body nonresonant processes and $\zcsm$ signal processes are scaled arbitrarily. The histogram of wrong-sign (WS) events is scaled by a factor of 1.18 to match the sideband data.}
\vspace{-0.5cm}
\label{fig:rmkds}
\end{figure}

Figure~\ref{fig:dsmass} shows the $\kaonp\kaonm\pim$ and $\kshort\kaonm$ invariant mass distributions for events at $\sqrt{s}=4.681\gev$, in which $\dsm$ peaks are clearly evident.
All combinations with invariant mass in the region $1.955<M(\kaonp\kaonm\pim)< 1.980\gevcc$ and $1.955<M(\kshort\kaonm)< 1.985\gevcc$ are identified as $\dsm$ meson candidates.
Figure~\ref{fig:rmkds} shows the $K^+\dsm$  recoil-mass spectrum for $\dsm$ candidate events at $\sqrt{s}=4.681\gev$, calculated using $RM(K^+\dsm)+M(\dsm)-m(\dsm)$.
Here, $RM(X)=||p_{\ee}-p_{X}||$, where $p_{\ee}$ is the four-momentum of the initial $\ee$ system and $p_X$ is the four-momentum of the system $X$, $M(\dsm)$ is the reconstructed $\dsm$ mass, and $m(\dsm)$ is the mass of the $\dsm$ reported by the PDG~\cite{pdg}.
The variable $RM(\kaonp\dsm)+M(\dsm)-m(\dsm)$  provides improved resolution compared to $RM(K^+D_s^-)$~\cite{Ablikim:2013emm1}.
A clear peak is seen in this distribution at the nominal $\dstzero$ mass, which corresponds to the final state $\kaonp \dsm\dstzero$.
There is also a contribution from $K^+\dsstm\dzero$, which appears as a broader structure beneath the $\kaonp \dsm\dstzero$ signal.
Therefore, we require $RM(\kaonp\dsm)+M(\dsm)-m(\dsm)$ to be in the interval $(1.990, 2.027) \gevcc$ to isolate the signal candidates of both signal processes.

To estimate the shape of combinatorial background, we use wrong-sign~(WS) combinations of $\dsm$ and $\kaonm$ candidates, rather than the right-sign $\dsm$ and $\kaonp$ candidates.
The WS $\kaonm\dsm$ recoil-mass distribution, scaled by a factor of 1.18, agrees with the data distribution in the sideband regions, (1.91, 1.95)$\gevcc$ and (2.08, 2.11)$\gevcc$, as shown in Fig.~\ref{fig:rmkds}.
The number of background events within the signal region is estimated to be {$282.6\pm12.0$} by a fit to the sideband data with a linear function, whose slope is determined from the WS data.
In addition, the WS events are used to represent the combinatorial-background distribution of the recoil mass of the bachelor $\kaonp$. This technique has been used previously in the observation of the $Z_c(4025)^+$ at BESIII~\cite{Ablikim:2013emm1}.
We validate the use of the WS data-driven background modeling of both the $RM(\kaonp\dsm)$ and $RM(\kaonp)$ spectra by comparing the corresponding distributions between WS combinations and background-only contributions. Furthermore, the $RM(\kaonp)$ distribution of the events in the sideband regions in Fig.~\ref{fig:rmkds} agrees well with that of the corresponding WS data.

\begin{figure}[tp!]
\centering
\vspace{-0.25cm}
\includegraphics[width=1.0\columnwidth]{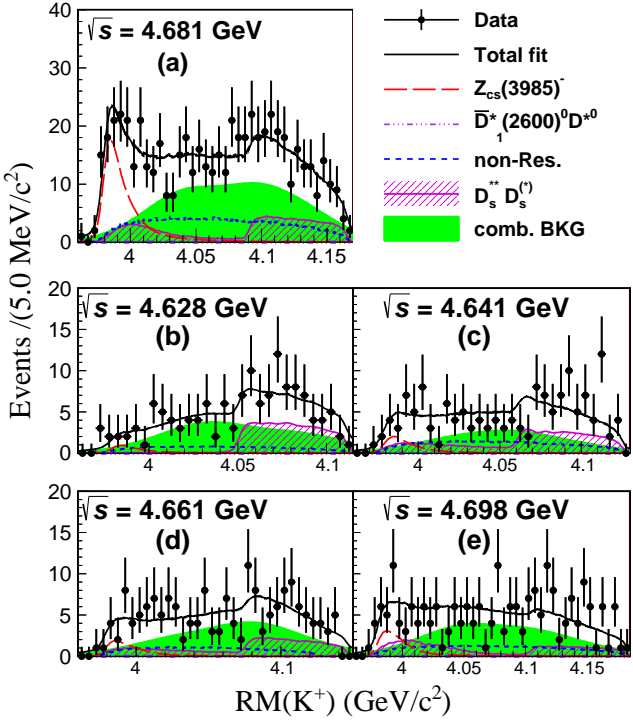}
\vspace{-0.5cm}
\caption{Simultaneous unbinned maximum likelihood fit to the $\kaonp$ recoil-mass spectra in data at $\sqrt{s}$=4.628, 4.641, 4.661, 4.681 and 4.698\gev. Note that the size of the $D^{*0}\bar{D}_1^{*}(2600)^0(\to D_s^- K^+)$ component is consistent with zero.}
\label{fig:rmkfit}
\end{figure}

\begin{figure}[tp]
\centering
\vspace{-0.25cm}
\includegraphics[width=0.65\columnwidth]{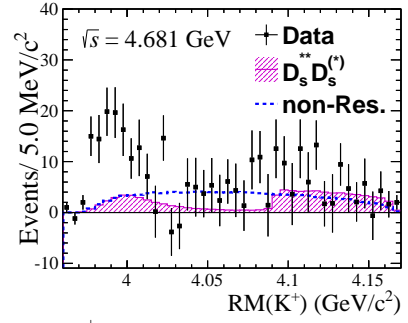}
\vspace{-0.5cm}
\caption{The $\kaonp$ recoil-mass spectrum in data at $\sqrt{s}=4.681\gev$ after subtraction of the combinatorial backgrounds. }
\label{fig:rmksub}
\end{figure}

Figure~\ref{fig:rmkfit}(a) shows the $RM(K^+)$ distribution for events at $\sqrt{s}=4.681\gev$; an enhancement is evident in the region $RM(\kaonp)<4\gevcc$ compared to the expectation from the WS events.
This is clearly illustrated in the $RM(K^+)$ distribution in data with subtraction of the WS component in Fig.~\ref{fig:rmksub}.
The enhancement cannot be attributed to the NR signal processes $\ee\to \kaonp(\dsm\dstzero+\dsstm\dzero)$.
To understand potential contributions from the processes $\ee\to D_s^{(*)-}D_s^{**+}(\to D^{(*)0}K^+)$ or $D^{(*)0}\bar{D}^{**0}(\to D_s^{(*)-}K^+)$, we examine all known $D_{(s)}^{**}$ excited states~\cite{pdg,Aaij:2019sqk} using MC simulation samples.
Dedicated exclusive MC studies show that none of these  processes, including possible interference effects, exhibit a narrow structure below 4.0$\gevcc$~\cite{supple}.

The following three processes that contain excited $D_{s}^{**+}$ background have potential contributions to the $RM(\kaonp)$ spectrum:
(1)~$\dsm D^{*}_{s1}(2536)^{+}(\to\dstzero\kaonp)$,
(2)~$\dsstm D^{*}_{s2}(2573)^{+}(\to\dzero\kaonp)$,
and (3)~$\dsm D^{*}_{s1}(2700)^{+}(\to\dstzero\kaonp)$.
We estimate their production cross sections by studying several control samples.
The yields for channel~(1) are estimated by analyzing the $D^{*}_{s1}(2536)^{+}$ peak in the $\dstzero\kaonp$ mass spectra using two separate partially reconstructed samples: $K^+\dsm$ (with $\dstzero$ missing) and $K^+\dstzero$ (with $\dsm$ missing).
For channel~(2), control samples are selected by reconstructing $\dzero\kaonp\gamma$ (with missing $\dsm$) or $\kaonp\dsstm$ (with missing $\dzero$).
The $D^{*}_{s2}(2573)^{+}$ yield is obtained from combined fits to the $\dzero\kaonp$ mass spectra. From this, the contribution from channel~(2) to the signal candidates in Fig.~\ref{fig:rmkfit} is evaluated.
For channel~(3), a control sample of $\ee\to \dsm D^{*}_{s1}(2700)^{+}(\to\dzero\kaonp)$ is selected by detecting the $\dsm\kaonp$ recoiling against a missing $\dzero$.
We then use the BF ratio of $\br{D^{*}_{s1}(2700)^{+}\to\dstzero\kaonp}/\br{D^{*}_{s1}(2700)^{+}\to\dzero\kaonp}=0.91\pm0.18$~\cite{Aubert:2009ah} to estimate the strength of this background contribution.
The shapes in $RM(\kaonp)$ of these three channels are extracted from MC samples, whereas the normalization is derived from the control samples.
The estimated background contributions of the channels~(1), (2) and (3) in the $RM(\kaonp)$ spectrum at $\sqrt{s}=4.681\gev$ are $54.4\pm8.0$, $19.1\pm7.6$ and $15.0\pm13.3$ events, respectively.
For the other energy points, the estimated yields of the three channels are given in Ref.~\cite{supple}.

Two processes with excited nonstrange $\bar{D}^{**0}$ states that produce potential enhancements around 4\,GeV/$c^2$ in $RM(\kaonp)$ are $D^{*0}\bar{D}_1^{*}(2600)^0(\to D_s^- K^+)$~\cite{pdg,Aaij:2019sqk} and $\dzero\bar{D}^{*}_{3}(2750)^{0}(\to D_s^{*-} K^+)$.
In these processes, the $RM(\kaonp)$ spectrum is distorted due to limited production phase space.
The first process is studied using an amplitude analysis of the control sample $\ee\to D^{*0}\bar{D}_1^{*}(2600)^0(\to D^- \pi^+)$ at all five energy points.
Since the ratio $\br{\bar{D}_1^{*}(2600)^0\to D_s^- K^+}/\br{\bar{D}_1^{*}(2600)^0\to D^- \pi^+}$ is unknown, it is difficult to project the results of the amplitude analysis into our signal channel.
Instead, we determine the ratio in our nominal fit, providing a constraint on the size of the $D^{*0}\bar{D}_1^{*}(2600)^0(\to D_s^- K^+)$ component at the different energy points.
For the second process, no significant signal is observed in the control sample $\ee\to\dzero\bar{D}^{*}_{3}(2750)^{0}(\to D^-\pi^+)$.
Assuming the relative BF ratio $\br{\bar{D}^{*}_{3}\to \dsstm\kaonp}/\br{\bar{D}^{*}_{3}\to D^-\pi^+}=4.1\%$~\cite{Godfrey:2015dva}, the contribution of the $\dzero{D}^{*}_{3}(2750)^{0}$ channel to Fig.~\ref{fig:rmkfit} is estimated to be $0.0\pm0.4$ events, and the corresponding upper limit is taken into account as a source of systematic uncertainty.

As no known processes explain the observed enhancement in the $RM(K^+)$ spectrum, which is very close to the threshold of $\dsm\dstzero~(3975.2\mevcc)$ and $\dsstm\dzero~(3977.0\mevcc)$,
we consider the possibility of describing the structure as a $\dsm\dstzero$ and $\dsstm\dzero$ resonance with a mass-dependent-width Breit-Wigner line shape, denoted as $\zcsm$.
A simultaneous unbinned maximum likelihood fit is performed to the $RM(K^+)$ spectra at all five energy points, as shown in Fig.~\ref{fig:rmkfit}.
The $\zcsm$ component is modeled by the product of an $S$-wave Breit-Wigner shape with a mass-dependent width of the following form
\begin{eqnarray}
\mathcal{F}_{j}(M) \propto \Big | \frac{ \sqrt{q \cdot p_{j}}}{M^2-m_{0}^2 + i m_{0}(f\Gamma_{1}(M)+(1-f)\Gamma_{2}(M))} \Big |^2,\nonumber
\end{eqnarray}
where
$\Gamma_{j}(M)=\Gamma_{0}\cdot\frac{p_{j}}{p_{j}^{*}}\cdot\frac{m_{0}}{M}$ with subscript $j=1$ and $j=2$ standing for the decays of $\zcsm\to\DDsone$ and $\zcsm\to\DDstwo$, respectively.
Here, $M$ is the reconstructed mass; $m_{0}$ is the resonance mass; $\Gamma_{0}$ is the width;
$q$ is the $\kaonp$ momentum in the initial $\ee$ system;
$p_{1}$ ($p_{2}$) is the $\dsm$ ($\dsstm$) momentum in the rest frame of the $\dsm\dstzero$ ($\dsstm\dzero$) system;
$p^*_{1}$ ($p^*_{2}$) is the $\dsm$ ($\dsstm$) momentum in the rest frame of the $\dsm\dstzero$ ($\dsstm\dzero$) system at $M$=$m_{0}$.
We define $f=[\mathcal{B}_1/(\mathcal{B}_1 + \mathcal{B}_2)]$,
where $\mathcal{B}_j$ is the BF of the $j$th decay.
We assume $f=0.5$ in the nominal fit and take variations of $f$ into account in the studies of systematic uncertainty.

The $Z_{cs}(3985)^{-}$ signal shape, which is used in the fit depicted in Fig.~\ref{fig:rmkfit}, is the $f$-dependent sum of the efficiency-weighted $\mathcal{F}_{j}$ functions convolved with a resolution function, which is obtained from MC simulation.
The resolution is about $5\mevcc$ and is asymmetric due to the contribution from ISR.
The parametrization of the combinatorial-background shape is derived from the kernel estimate~\cite{Cranmer:2000du} of the WS distribution, whose normalization is fixed to the number of the fitted background events within the decorrelated $RM(K^+D_s^-)$ signal window.
The shapes of the NR and $D^{*0}\bar{D}_1^{*}(2600)^0(\to D_s^- K^+)$ signals are taken from the MC simulation.
The size of the NR component at each energy point and the ratio $\br{\bar{D}_1^{*}(2600)^0\to D_s^- K^+}/\br{\bar{D}_1^{*}(2600)^0\to D^- \pi^+}$ are free parameters in the fit.
In addition, a component that describes the total contributions of the excited  $D_{s}^{**+}$  processes is included, whose shape is taken from MC simulation and its size is fixed according to the yields estimated from the control-sample studies.

From the fit, the parameters $m_{0}$ and $\Gamma_{0}$
are determined to be
$(3985.2^{+2.1}_{-2.0})\mevcc$ and $(13.8^{+8.1}_{-5.2})\mev$, respectively.
The significance of the signal is calculated taking into account the look-elsewhere effect~\cite{Gross:2010qma}, where 5000 pseudodatasets are produced with the sum of null-$\zcsm$ models and fitted with the same strategy as the nominal fit to obtain the distribution of $-2\ln (L_0/L_{\rm max})$, where $L_0$ and $L_{\rm max}$ are fitted likelihood values under the null-$\zcsm$ hypothesis and alternative hypothesis, respectively.
In the generation of the pseudodata, the systematic uncertainties relevant to determine the signal yields, as marked in Table~II in Ref.~\cite{supple} are considered.
The resulting distribution is found to be well described by a $\chi^2$ distribution with 13.8 degrees of freedom. With an observed value of $-2\ln (L_0/L_{\rm max})=59.14$, we obtain a significance of $5.3\,\sigma$.
The number of $\zcsm$ events observed at $\sqrt{s}=4.681\gev$ is the most prominent compared to the other four energy points.
If we fit only to data at $\sqrt{s}=4.681\gev$, we obtain consistent $\zcsm$ resonance parameters.

The Born cross section $\sigma^{B}[\ee\to \kaonp\zcsm+ {\rm c.c.}]$ times the sum of BFs of the decays $\zcsm\to D_s^-D^{*0}+D_s^{*-}D^{0}$ is equal to
$\frac{n_{\rm sig}}{\mathcal{L}_{\rm int}f_{\rm corr}\bar{\varepsilon}}$, where $n_{\rm sig}$ is the number of the observed signal events, $\mathcal{L}_{\rm int}$ is the integrated luminosity, and $\bar{\varepsilon}$ is the BF-weighted detection efficiency.
We define $f_{\rm corr}\equiv(1+\delta_{\rm ISR}) \frac{1}{|1-\Pi|^2}$, where $(1+ \delta_{\rm ISR})$ is the radiative-correction factor and $\frac{1}{|1-\Pi|^2}$ is the vacuum-polarization factor~\cite{Jegerlehner:2011mw}. The numerical results are listed in Table~\ref{tab:xs}.

\begin{table}[tp]
  \begin{center}
  \footnotesize
\caption{The results for the cross section measurement at each energy point. The upper limits in the parenthesis correspond to 90\% confidence level after considering the systematic uncertainties.}
  \begin{tabular}{cccccc}
      \hline \hline
       $\sqrt{s}(\gev)$ &  $\mathcal{L}_{\rm int}$($\invpb$)  & $n_{\rm sig}$ & $f_\mathrm{corr}\bar{\varepsilon}(\%)$ & $\sigma^B\cdot\mathcal{B}\;$(pb) \\ \hline
 $4.628$ & 511.1  	&$4.2^{+6.1}_{-4.2}$    & $1.03$   &$0.8^{+1.2}_{-0.8}\pm0.6\,(<3.0)$ \\
 $4.641$ & 541.4    &$9.3^{+7.3}_{-6.2}$    & $1.09$   &$1.6^{+1.2}_{-1.1}\pm1.3\,(<4.4)$ \\
 $4.661$ & 523.6    &$10.6^{+8.9}_{-7.4}$   & $1.28$   &$1.6^{+1.3}_{-1.1}\pm0.8\,(<4.0)$ \\
 $4.681$ & 1643.4  	&$85.2^{+17.6}_{-15.6}$ & $1.18$   &$4.4^{+0.9}_{-0.8}\pm1.4$         \\
 $4.698$ & 526.2    &$17.8^{+8.1}_{-7.2}$   & $1.42$   &$2.4^{+1.1}_{-1.0}\pm1.2\,(<4.7)$ \\ \hline \hline
    \end{tabular}
    \label{tab:xs}
  \end{center}
  \end{table}

Sources of systematic uncertainties on the measurement of the $\zcsm$ resonance parameters and the cross section are studied, as explained in Ref.~\cite{supple}.
The main sources include the mass scaling, detector resolution, the signal model, background models and the input cross section line shape for $\sigma^{B}[\ee\to \kaonp\zcsm]$.
The contributions to the systematic uncertainties on the resonance parameters and cross sections are give in Table~\ref{tab:syst1} and Ref.~\cite{supple}, respectively.
In addition, the global signal significances after taking into account the look-elsewhere effect under different systematic effects are listed in Table~\ref{tab:syst1}.

\begin{table}[tp]
  \begin{center}
  \footnotesize
   \caption{Summary of systematic uncertainties on the $\zcsm$ resonance parameters. The total systematic uncertainty corresponds to a quadrature sum of all individual items.
 The global signal significance after taking into account the systematic item marked with $*$ is listed.}
  \begin{tabular}{lccc}
      \hline \hline
      Source                    & Mass($\mevcc$)  & Width($\mev$)  & Significance\\ \hline
      Mass  scale               & 0.5             &                & \\
      Resolution$^*$            & 0.2             & 1.0            & 5.7$\,\sigma$\\
      $f$ factor$^*$            & 0.2             & 1.0            & 5.6$\,\sigma$\\
      Signal model$^*$          & 1.0             & 2.6            & 5.7$\,\sigma$\\
      Backgrounds$^*$           & 0.5             & 0.5            & 5.6$\,\sigma$\\
      Efficiencies              & 0.1             & 0.2            & \\
      $D_{(s)}^{**}$ states$^*$ & 1.0             & 3.4            & 5.4$\,\sigma$\\
      $\sigma^B[\kaonp\zcsm]$   & 0.6             & 1.7            & \\
          \hline
          Total                 & 1.7             & 4.9            &\\
        \hline\hline
    \end{tabular}
    \label{tab:syst1}
  \end{center}
  \end{table}

In summary, we study the reactions $\ee \to \kaonp(\dsm\dstzero+\dsstm\dzero)$ based on 3.7\,fb$^{-1}$ of data collected at $\sqrt{s}=$4.628, 4.641, 4.661, 4.681, and 4.698$\gev$, and observe an enhancement near the $\dsm\dstzero$ and $\dsstm\dzero$ mass thresholds in the $K^{+}$ recoil-mass spectrum for events collected at $\sqrt{s}=4.681\gev$.
While the known charmed mesons cannot explain the excess, it matches a hypothesis of a $\dsm\dstzero$ and $\dsstm\dzero$ resonant structure $\zcsm$ with a mass-dependent-width Breit-Wigner line shape well; a fit gives the resonance mass
of $(3985.2^{+2.1}_{-2.0}\pm1.7)\mevcc$ and width of $(13.8^{+8.1}_{-5.2}\pm4.9)\mev$.
This corresponds to a pole position
$m_{\rm pole}-i\frac{\Gamma_{\rm pole}}{2}$ of
\begin{eqnarray}
   m_{\rm pole}[\zcsm]&=&(3982.5^{+1.8}_{-2.6}\pm2.1)\mevcc, \nonumber\\
  \Gamma_{\rm pole}[\zcsm]&=&(12.8^{+5.3}_{-4.4}\pm3.0)\mev. \nonumber
\end{eqnarray}
The first uncertainties are statistical and the second are systematic.
The significance of this resonance hypothesis is estimated to be $5.3\,\sigma$ over the pure contributions from the conventional charmed mesons.
The $\zcsm$ candidate reported here would couple to at least one of $\dsm\dstzero$ and $\dsstm\dzero$, and has unit charge, the quark composition is most likely $c\bar{c}s\bar{u}$.
Hence, it would become the first $Z_{cs}$ tetraquark candidate observed.
The measured mass is close to the mass threshold of $D_s\bar{D}^*$ and $D_s^*\bar{D}$, which is consistent with the theoretical calculations in Ref.~\cite{Ebert:2008kb, Lee:2008uy,Dias:2013qga,Chen:2013wca}.
In addition, the Born cross sections $\sigma^B[e^+e^-\to K^+Z_{cs}(3985)^{-}+ {\rm c.c.}]$ times the sum of the branching fractions for $\zcsm\to D_s^-D^{*0}+D_s^{*-}D^{0}$ decays are measured at the five energy points.
Because of the limited size of the statistics, only a one-dimensional fit is implemented and the potential interference effects are neglected.
As shown in Figs.~5 and 6 of Ref.~\cite{supple}, we find no evidence for enhancements due to interference below 4 GeV/$c^2$.
Even so, the properties of the observed excess might not be fully explored and there exist other possibilities of explaining the near-threshold enhancement.
To further improve studies of the excess, more statistics are necessary in order to carry out an amplitude analysis.


The BESIII Collaboration thanks the staff of BEPCII and the IHEP computing center for their strong support. This work is supported in part by National Key Basic Research Program of China under Contract Nos. 2015CB856700; National Key Research and Development Program of China under Nos. 2020YFA0406300, 2020YFA0406400; National Natural Science Foundation of China (NSFC) under Contracts Nos. 11625523, 11635010, 11735014, 11805086, 11822506, 11835012, 11935015, 11935016, 11935018, 11961141012; the Chinese Academy of Sciences (CAS) Large-Scale Scientific Facility Program; the CAS PIFI program; Joint Large-Scale Scientific Facility Funds of the NSFC and CAS under Contracts Nos. U1732263, U1832207; CAS Key Research Program of Frontier Sciences under Contracts Nos. QYZDJ-SSW-SLH003, QYZDJ-SSW-SLH040; 100 Talents Program of CAS; the Fundamental Research Funds for the Central Universities; INPAC and Shanghai Key Laboratory for Particle Physics and Cosmology; ERC under Contract No. 758462; German Research Foundation DFG under Contracts Nos. 443159800, Collaborative Research Center CRC 1044, FOR 2359, GRK 2149; Istituto Nazionale di Fisica Nucleare, Italy; Ministry of Development of Turkey under Contract No. DPT2006K-120470; National Science and Technology fund; Olle Engkvist Foundation under Contract No. 200-0605; STFC (United Kingdom); The Knut and Alice Wallenberg Foundation (Sweden) under Contract No. 2016.0157; The Royal Society, UK under Contracts Nos. DH140054, DH160214; The Swedish Research Council; U. S. Department of Energy under Contracts Nos. DE-FG02-05ER41374, DE-SC-0012069.


\onecolumngrid

\newpage
\appendix
\setcounter{table}{0}
\setcounter{figure}{0}

{\large \bf Supplemental Material for ``Observation of a Near-Threshold Structure in the $K^+$ Recoil Mass Spectra in $e^+e^-\to K^+(D_s^- D^{*0}+D^{*-}_s D^0)$'' }

\begin{appendices}
\section{Additional information: studies of the excess in $\kaonp$ recoil-mass spectrum}

\begin{figure*}[htp]
\centering
\subfigure[Recoil mass of $\kaonp\dsm$ at $\sqrt{s}=4.628\gev$.]
{\includegraphics[width=0.35\textwidth]{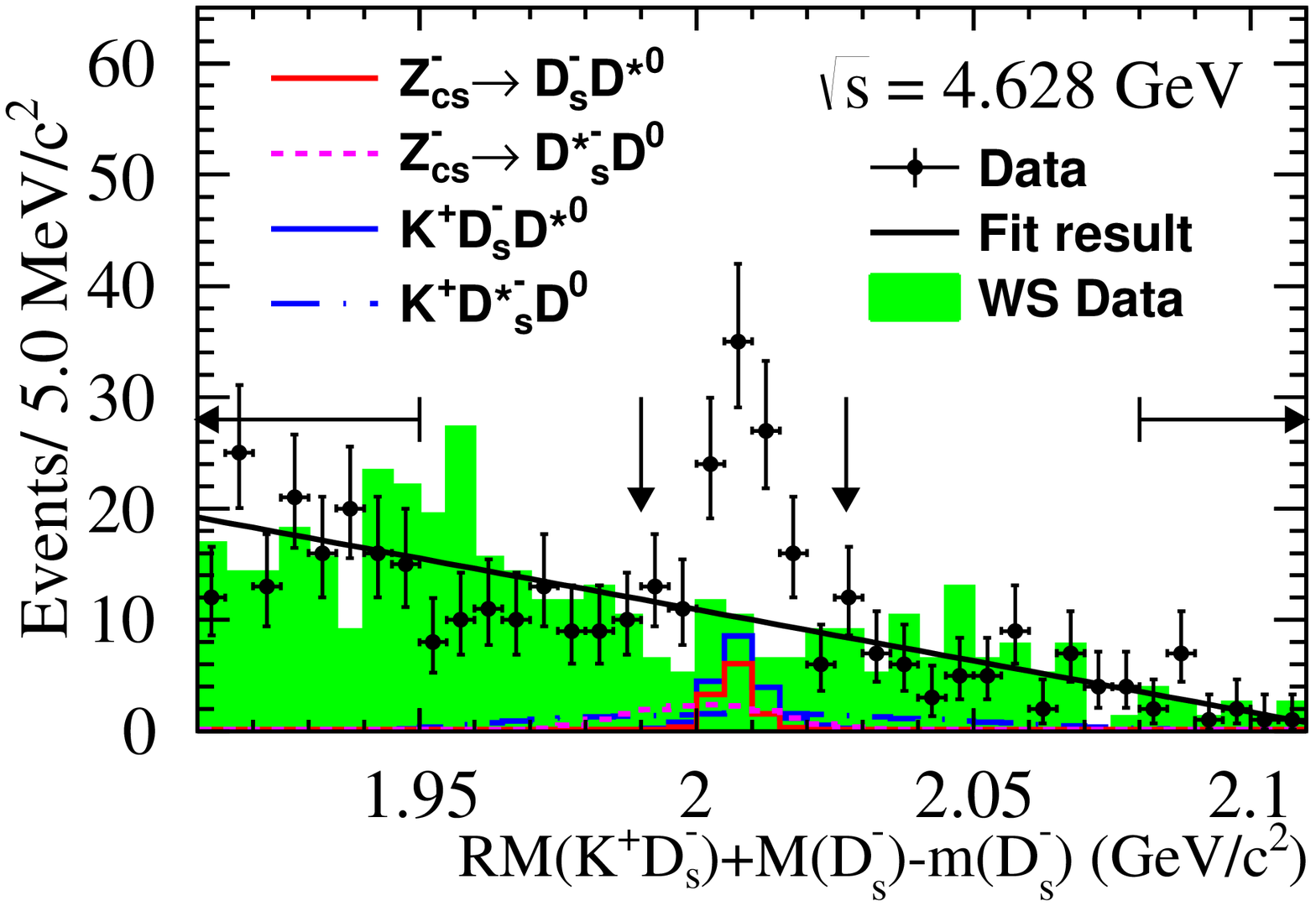} \label{fig:fit_RMKD_4626} }
\hspace{0.1\linewidth}
\subfigure[Recoil mass of $\kaonp\dsm$ at $\sqrt{s}=4.641\gev$.]
{\includegraphics[width=0.35\textwidth]{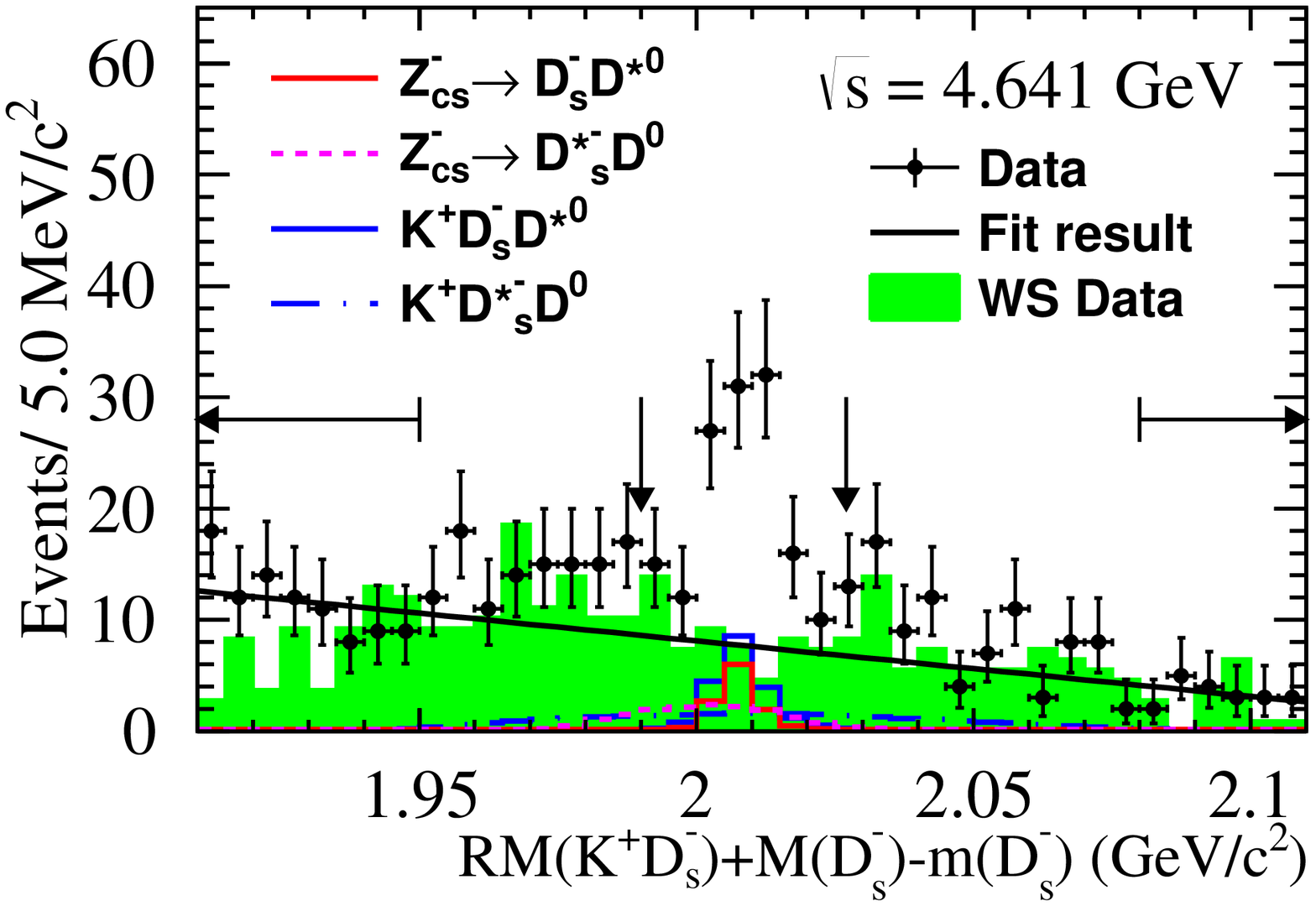} \label{fig:fit_RMKD_4640} }
\subfigure[Recoil mass of $\kaonp\dsm$ at $\sqrt{s}=4.661\gev$.]
{\includegraphics[width=0.35\textwidth]{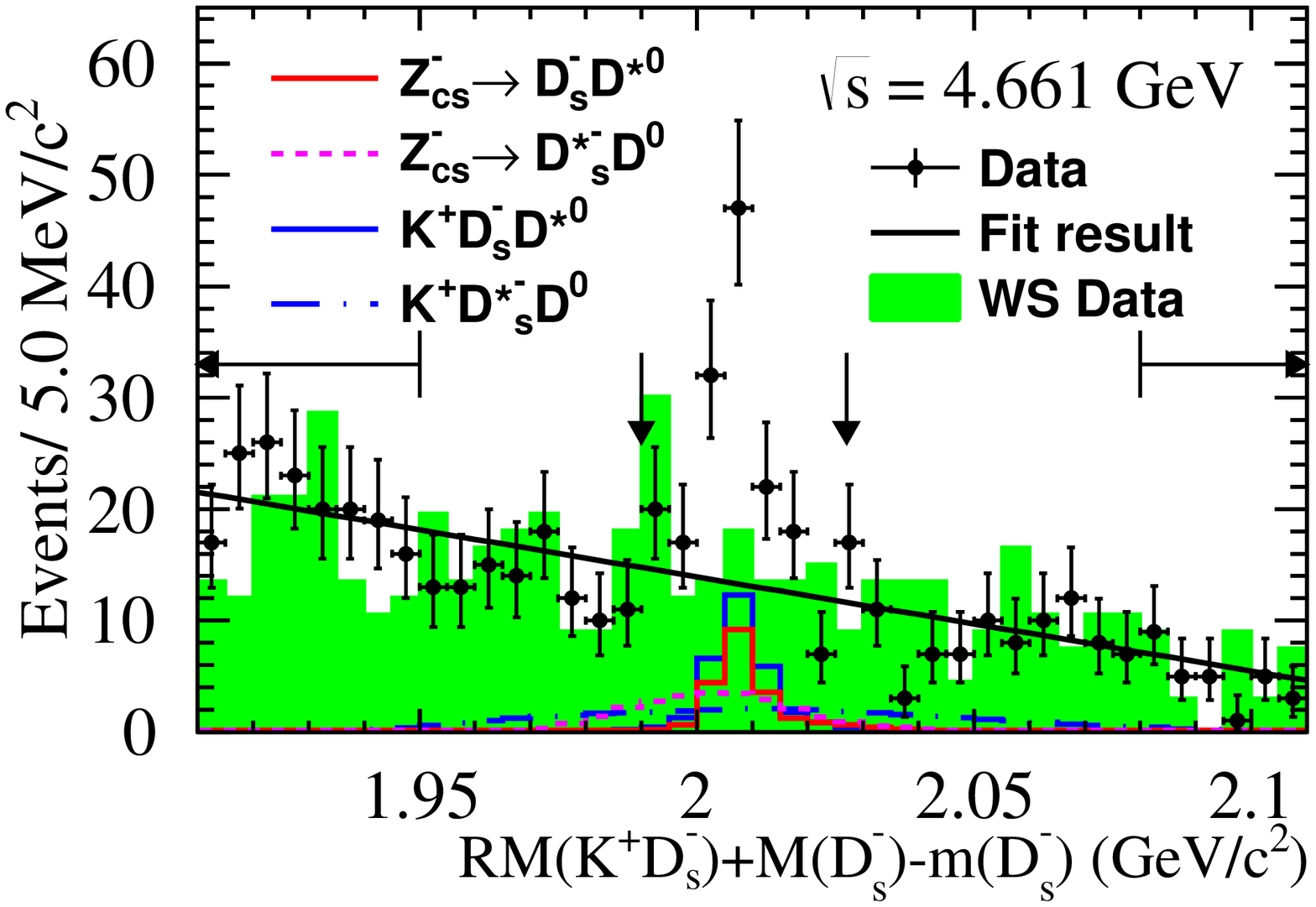} \label{fig:fit_RMKD_4660} }
\hspace{0.1\linewidth}
\subfigure[Recoil mass of $\kaonp\dsm$ at $\sqrt{s}=4.698\gev$.]
{\includegraphics[width=0.35\textwidth]{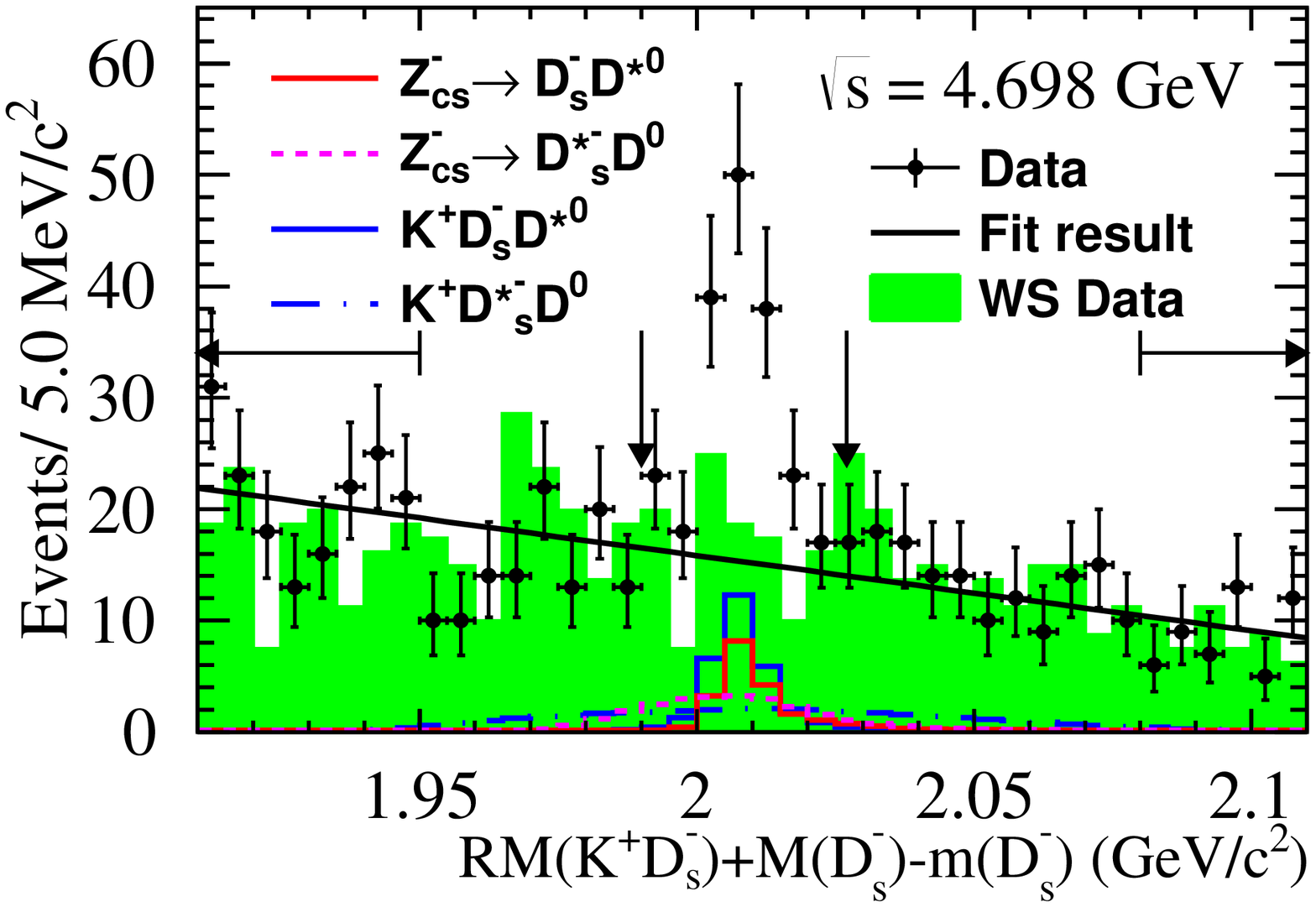} \label{fig:fit_RMKD_4700} }
\caption{Distribution of the $\kaonp\dsm$ recoil-mass in data and signal MC samples at different center-of-mass energies. Definitions of plotted components are the same as those in Fig.~2 of the main paper.}
\label{fig:fit_RMKDothers}
\end{figure*}

\begin{figure}[h]
\centering
\includegraphics[width=0.4\linewidth]{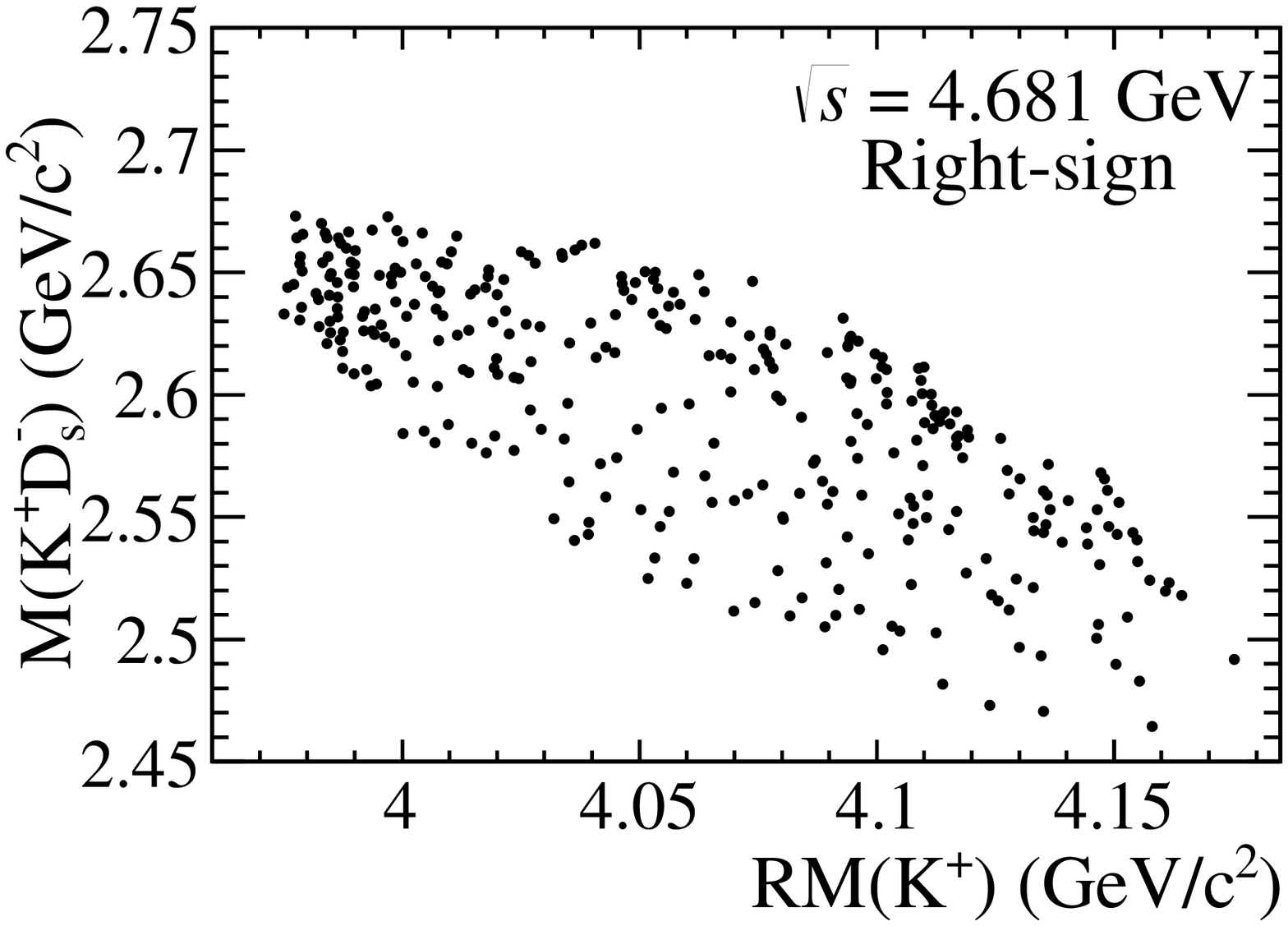}
\hspace{0.1\linewidth}
\includegraphics[width=0.4\linewidth]{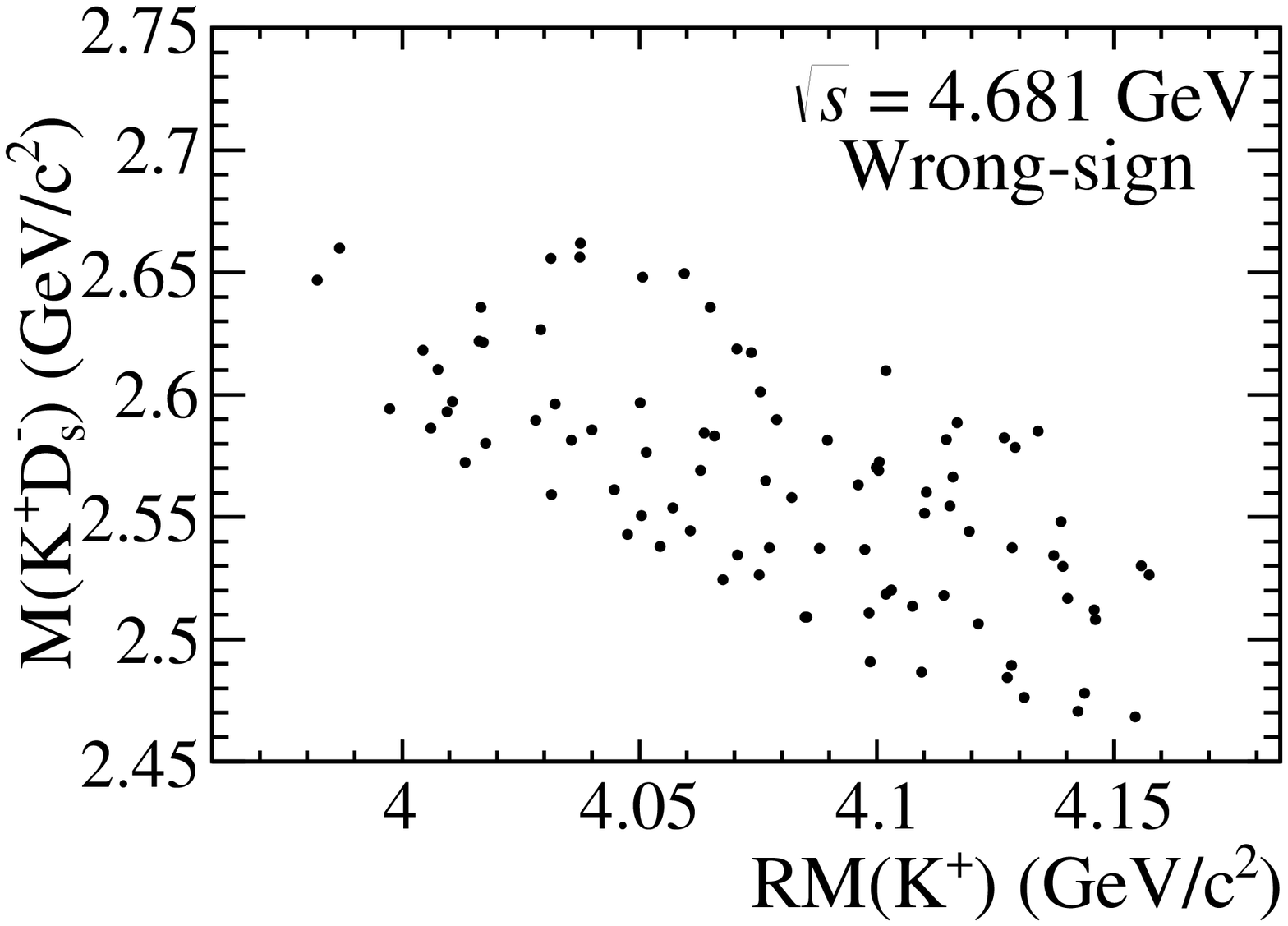}
\vspace{-0.25cm}
\caption{Two-dimensional distributions of $M(K^+\dsm)$ vs. $RM(\kaonp)$ for data in the signal region (left) and WS events (right) at $\sqrt{s}=4.681\gev$.}
\vspace{-0.25cm}
\label{fig:2dplot}
\end{figure}

Figure~\ref{fig:fit_RMKDothers} shows the distribution of the $\kaonp\dsm$ recoil-mass in data and MC simulation samples at $\sqrt{s}=4.628$, 4.641, 4.661 and $4.698\gev$, after the same selection criteria as those imposed for the data shown in Fig.~2 of the main letter. Table~\ref{tab:DDh_size} lists the estimated sizes of excited $D_{s}^{**+}$ or $\bar{D}^{**0}$ contributions at each energy point, quoted in the simultaneous fit. In addition, two-dimensional plots of $M(K^+\dsm)$ versus $RM(\kaonp)$ in data for events in the signal region and WS events at $\sqrt{s}=4.681\gev$  are shown in Fig.~\ref{fig:2dplot}.

\begin{table}[hp!]
  \begin{center}
\caption{Summary of the estimated sizes of excited $D_{s}^{**+}$ or $\bar{D}^{**0}$ contributions at each energy point. ``$-$'' means the production is not allowed kinematically. }
  \begin{tabular}{c|ccccc}
      \hline \hline
$\sqrt{s}(\gev)$  & $4.628$ & $4.641$ & $4.661$ & $4.681$ & $4.698$ \\ \hline
$D_{s1}(2536)^{+}(\kaonp\dstzero)\dsm$              &$41.2\pm6.3$ &$26.2\pm5.4$ &$23.9\pm5.6$ &$54.4\pm8.0$  &$15.3\pm4.2$    \\   \hline
$D^{*}_{s2}(2573)^{+}(\kaonp\dzero)\dsstm$          &$-$          &$-$          &$-$          &$19.1\pm7.6$  &$17.3\pm7.3$    \\   \hline
$D^{*}_{s1}(2700)^{+}(\kaonp\dstzero)\dsm$          &$0.0\pm1.8$  &$18.6\pm8.7$ &$16.6\pm7.8$ &$15.0\pm13.3$ &$7.7\pm8.4$    \\   \hline
$\bar{D}^{*}_{3}(2750)^{0}(\to\dsstm\kaonp)\dzero$  &$0.0\pm0.1$  &$0.0\pm0.2$  &$0.0\pm0.2$  &$0.0\pm0.4$   &$0.0\pm0.5$    \\   \hline
    \end{tabular}
    \label{tab:DDh_size}
  \end{center}
  \end{table}

\clearpage
\section{Exploration of  potential $D_{(s)}^{**}$ backgrounds}

\begin{figure*}[hpt]
\centering
\subfigure[$D_{s1}(2536)^{+}(\to\dstzero\kaonp)\dsm$]
{\includegraphics[width=0.3\textwidth]{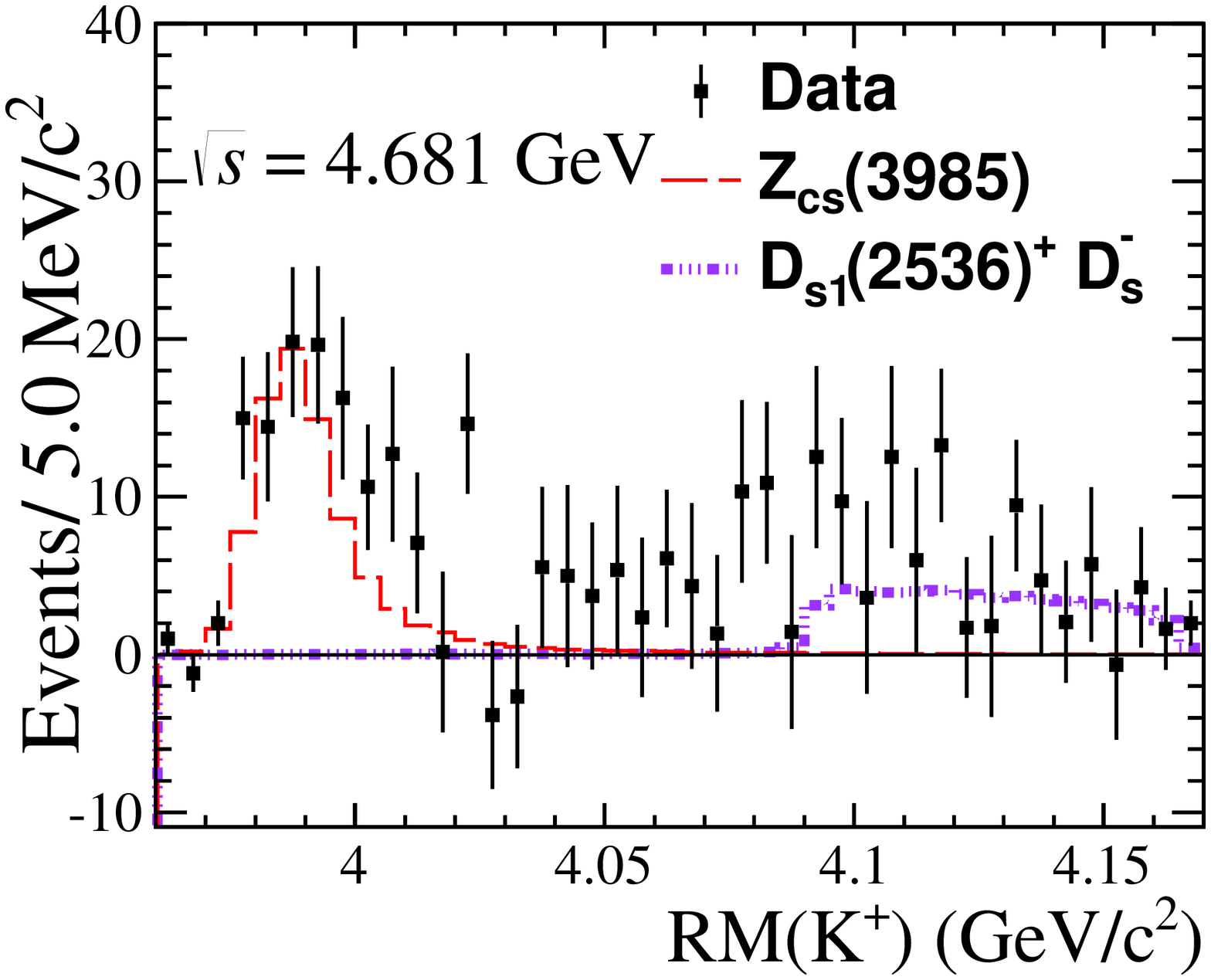}}
\subfigure[$D^{*}_{s2}(2573)^{+}(\to\dzero\kaonp)\dsstm$]
{\includegraphics[width=0.3\textwidth]{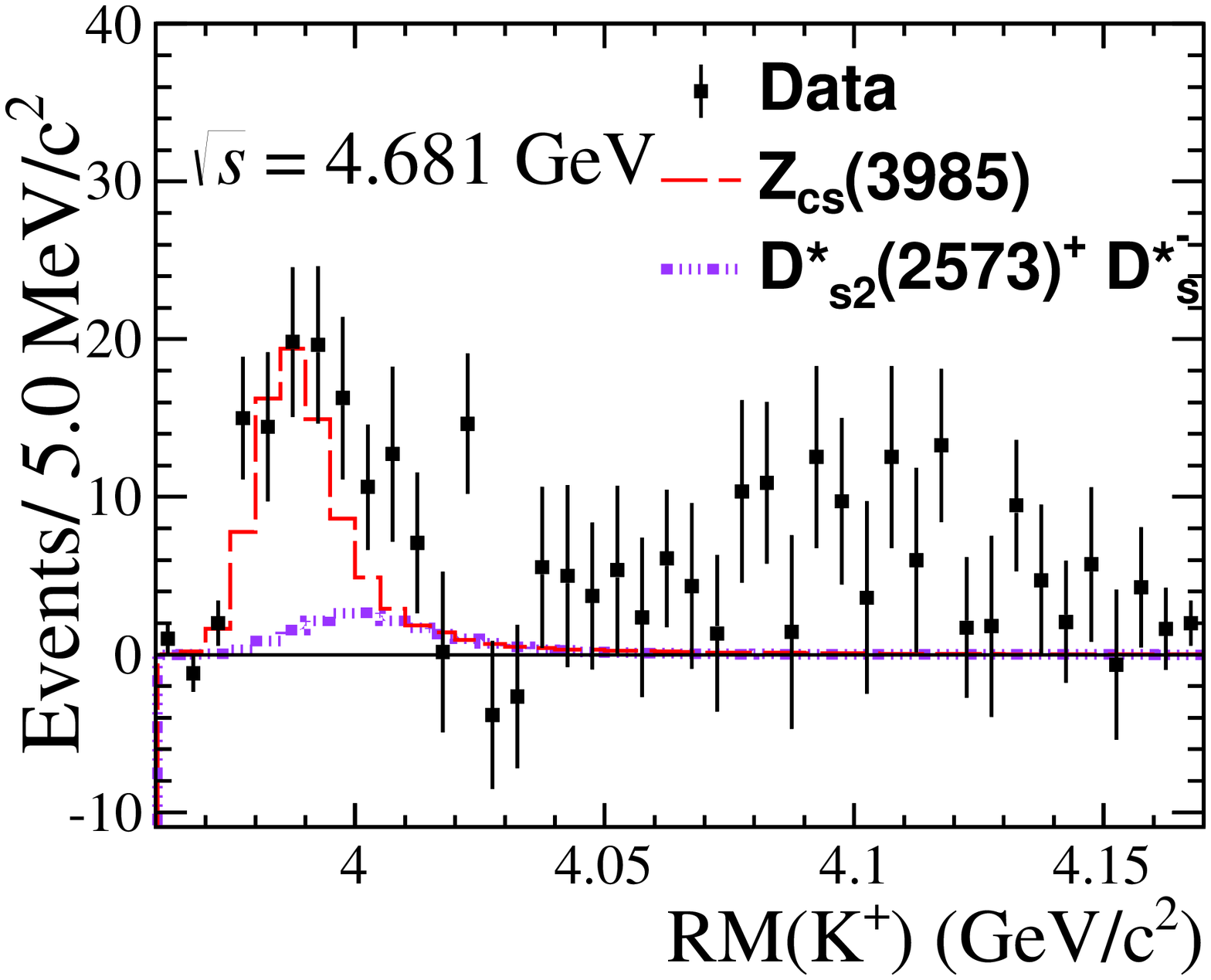}}
\subfigure[$D^{*}_{s1}(2700)^{+}(\to\dstzero\kaonp)\dsm$]
{\includegraphics[width=0.3\textwidth]{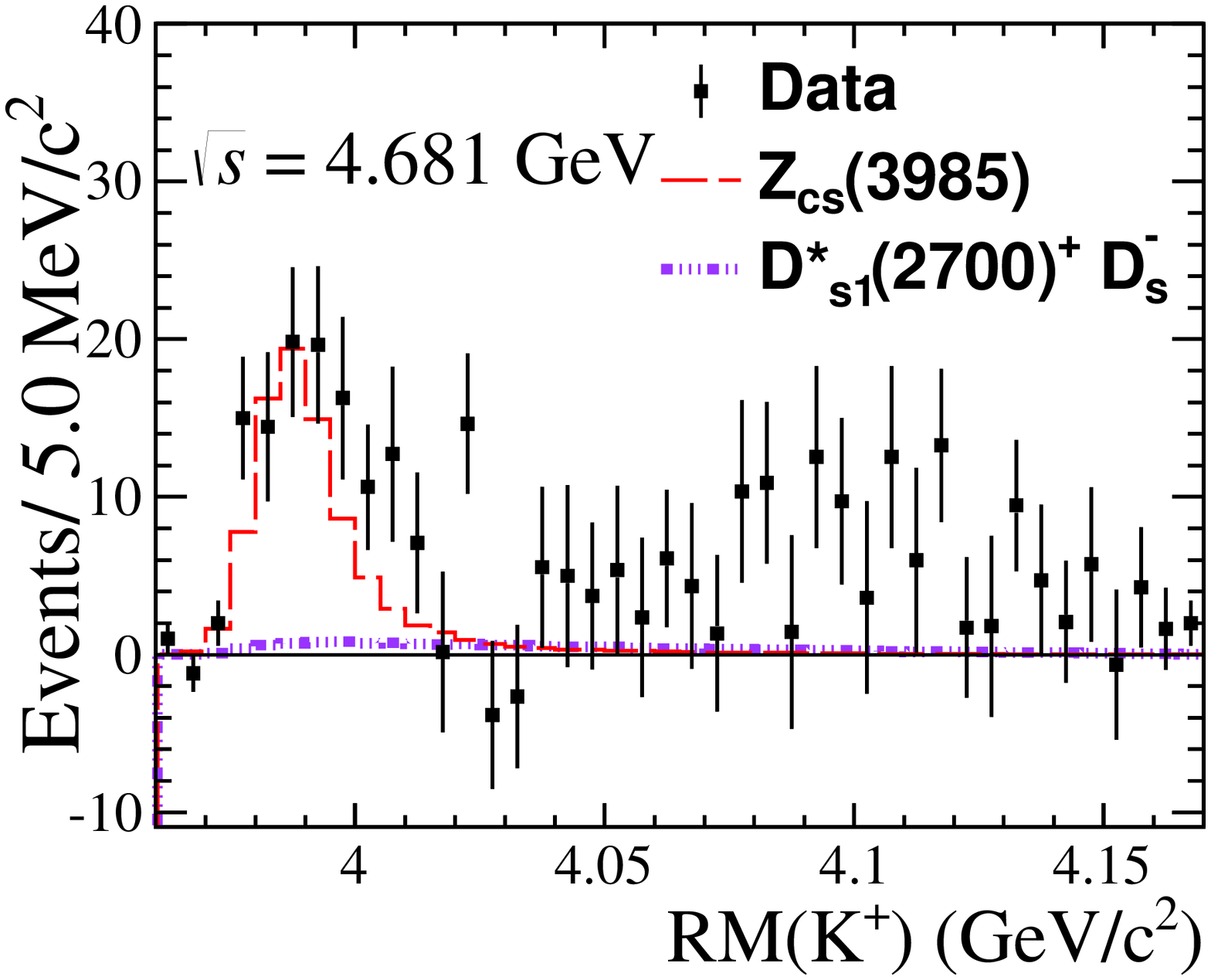}}
\caption{ $\kaonp$ recoil-mass spectra in data with the WS background contributions subtracted, and MC simulations of the  excited $D_s^{**}$ states in $\ee\to D_{s}^{**+}D_{s}^{(*)-}$. The $\zcsm$ shapes are normalized to the yields observed in data and those of the $D_s^{**}$ states are scaled according to the control samples.}
\label{fig:DDh_RMK_Ds}
\end{figure*}

\begin{figure*}[hpt]
\centering
\subfigure[$\bar{D}^{*}_{2}(2460)^{0}(\to\dsm\kaonp)\dstzero$]
{\includegraphics[width=0.3\textwidth]{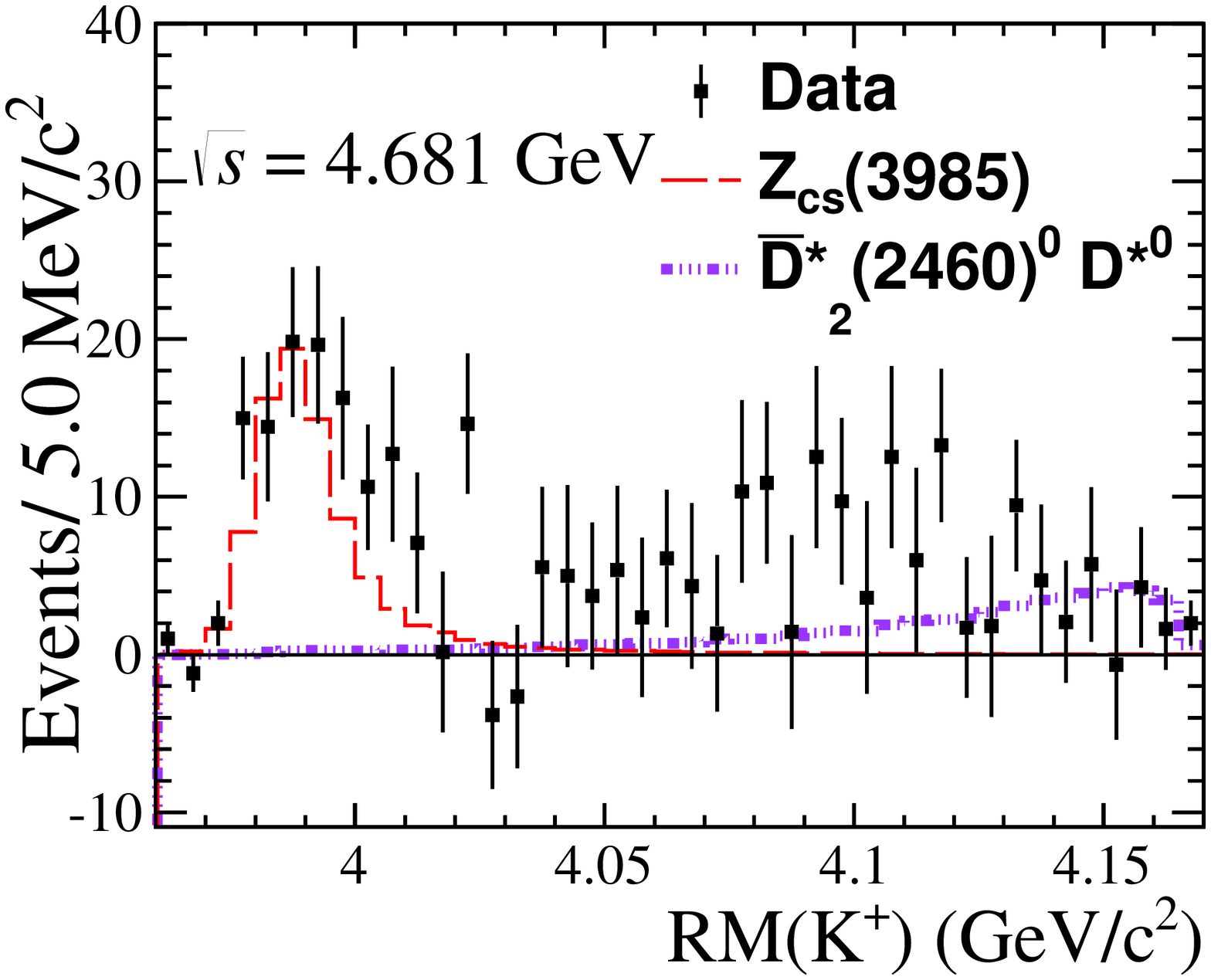}}
\subfigure[$\bar{D}(2550)^{0}(\to\dsstm\kaonp)\dzero$]
{\includegraphics[width=0.3\textwidth]{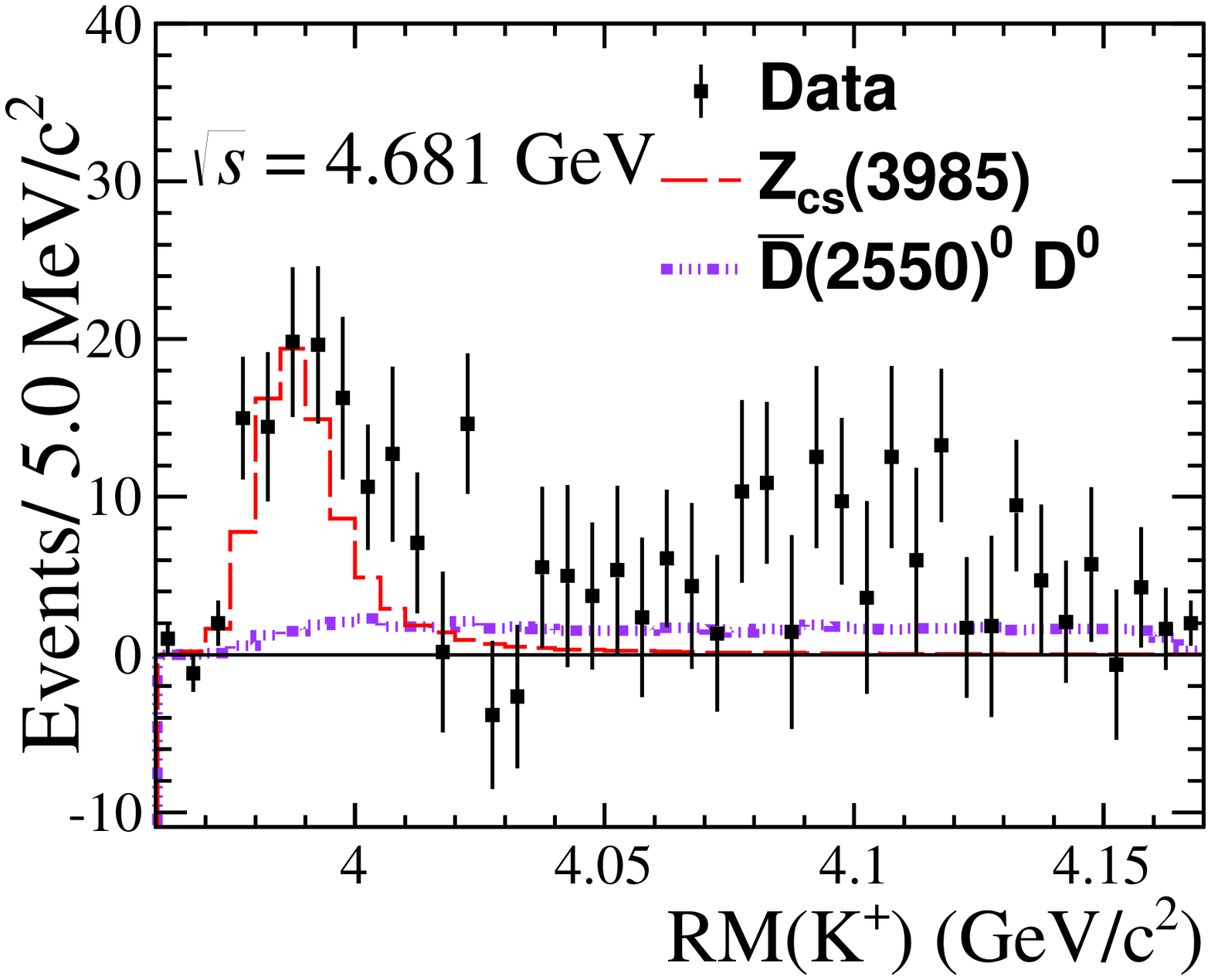}}
\subfigure[$\bar{D}^{*}_{1}(2600)^{0}(\to\dsm\kaonp)\dstzero$]
{\includegraphics[width=0.3\textwidth]{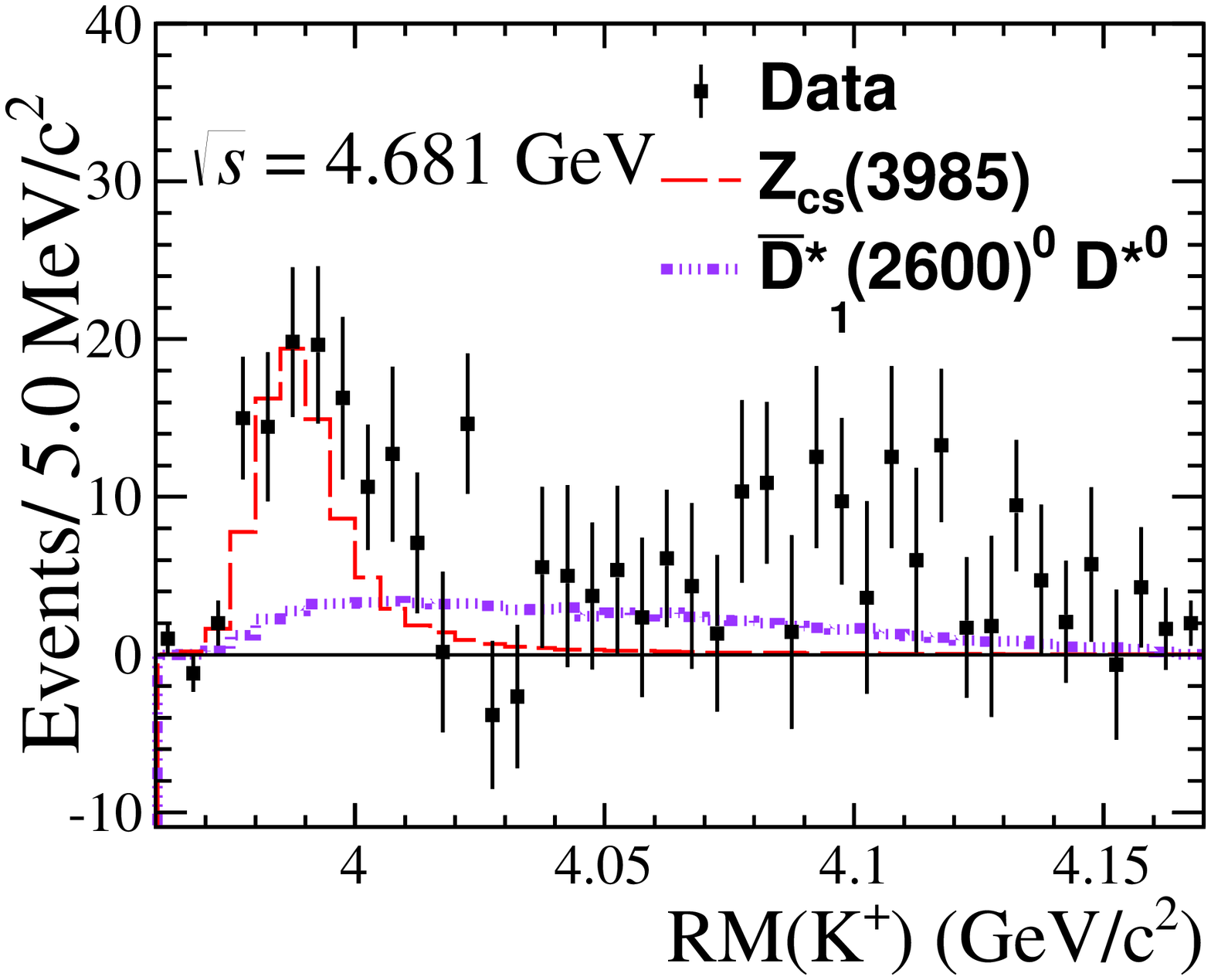}}
\subfigure[$\bar{D}^{*}_{1}(2600)^{0}(\to\dsstm\kaonp)\dzero$]
{\includegraphics[width=0.3\textwidth]{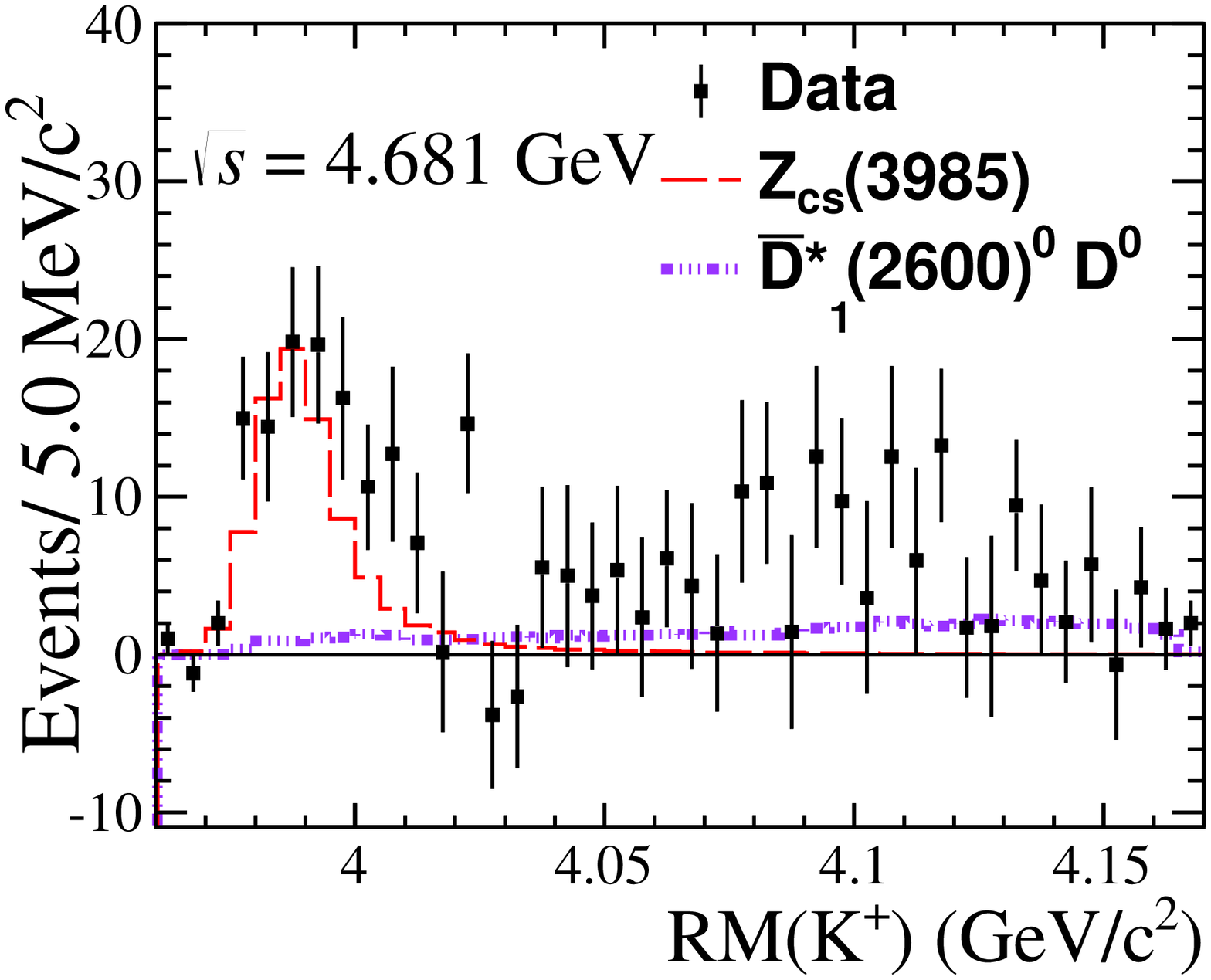}}
\subfigure[$\bar{D}(2740)^{0}(\to\dsstm\kaonp)\dzero$]
{\includegraphics[width=0.3\textwidth]{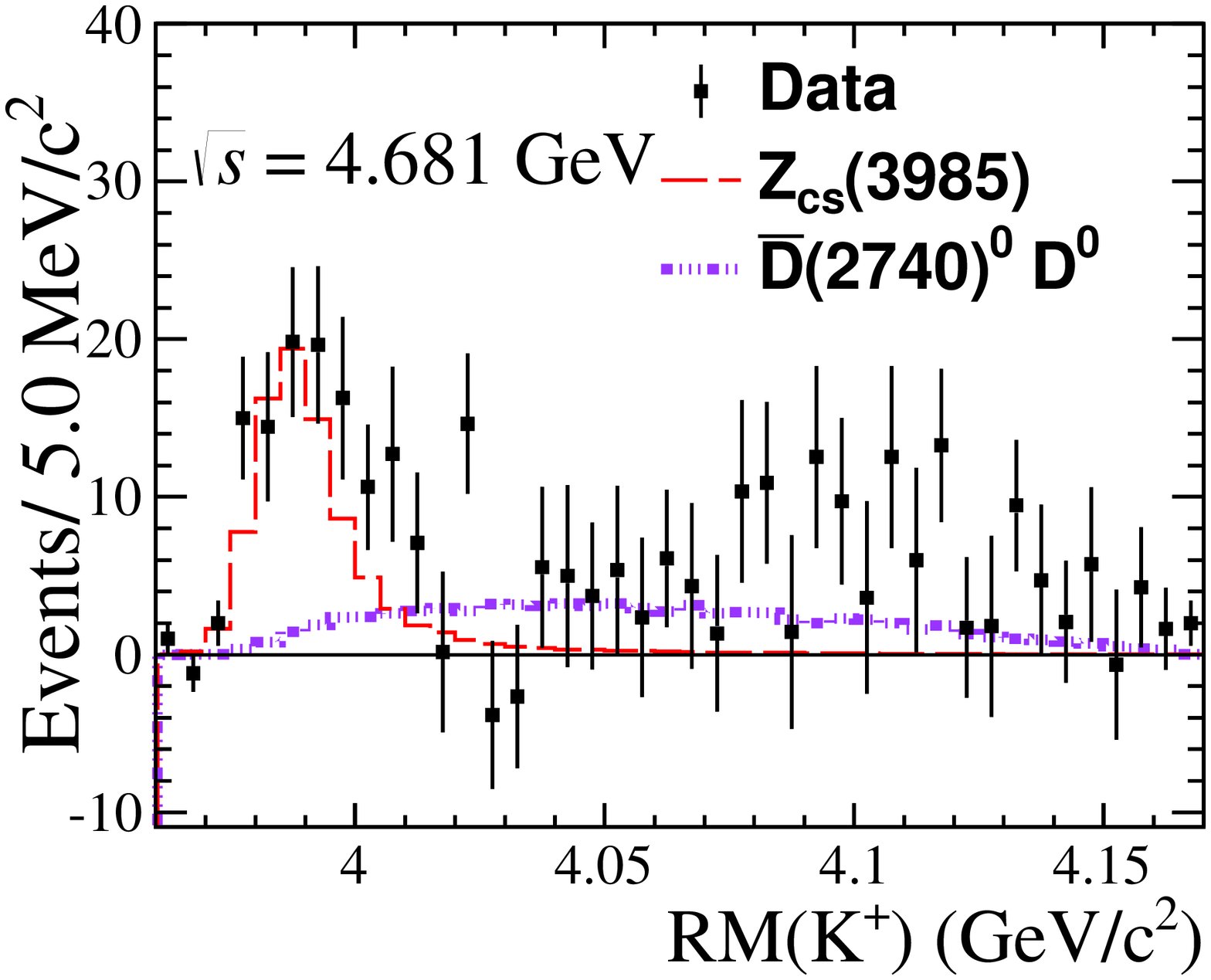}}
\subfigure[$\bar{D}^{*}_{3}(2750)^{0}(\to\dsstm\kaonp)\dzero$]
{\includegraphics[width=0.3\textwidth]{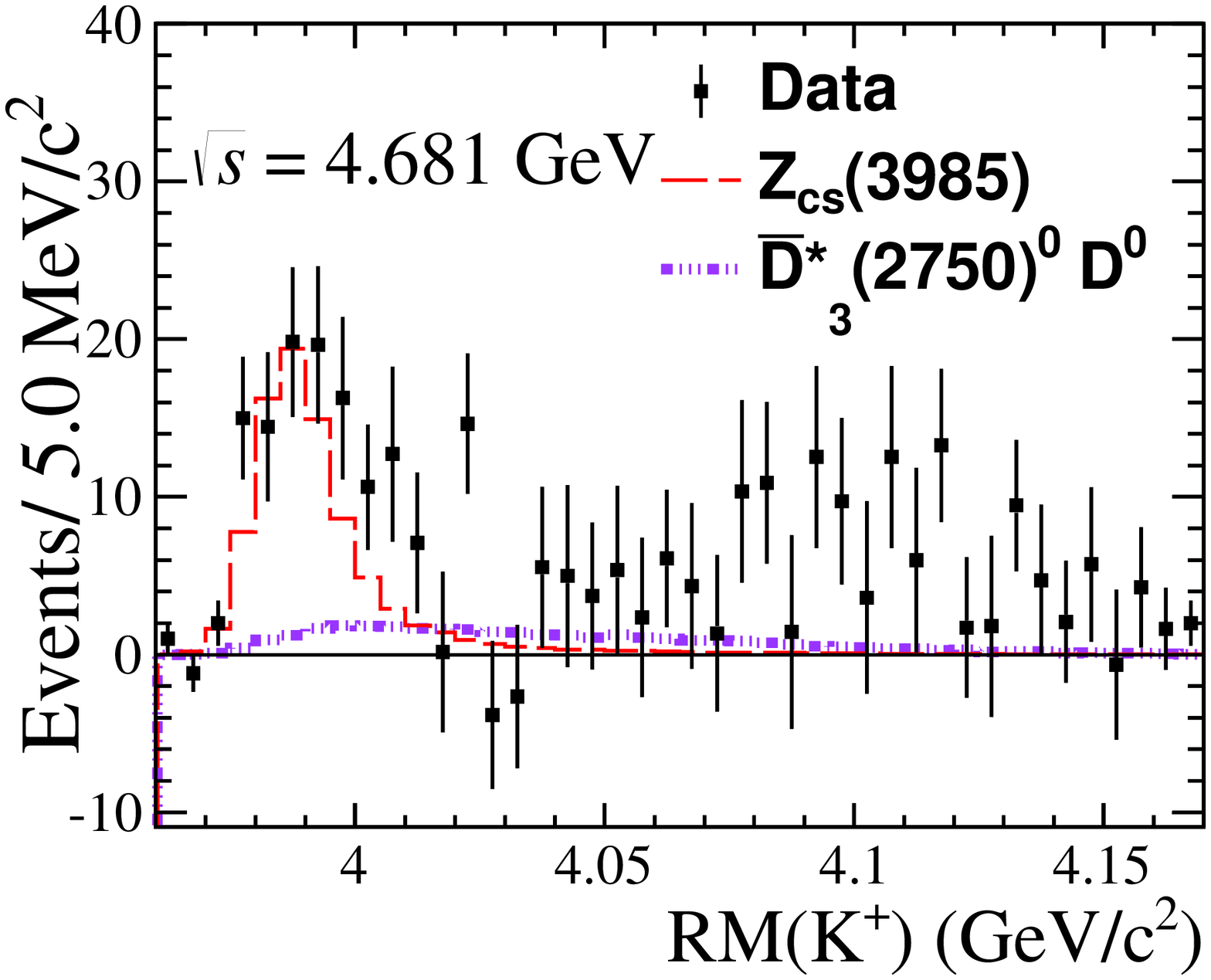}}
\caption{ $\kaonp$ recoil-mass spectra in data with the WS background contributions subtracted, and MC simulations of the excited $D^{**}$ states in  $\ee\to \bar{D}^{**0}D^{(*)0}$. The $\zcsm$ shape is normalized to the yields observed in data and the shape of the $\bar{D}^{**0}$ states is arbitrarily scaled.}
\label{fig:DDh_RMK_D}
\end{figure*}

\begin{figure*}[hpt]
\centering
\subfigure[\footnotesize $\bar{D}(2550)^{0}\dzero$ and $\bar{D}_{1}^{*}(2600)^{0}\dzero$ ]
{\includegraphics[width=0.3\textwidth]{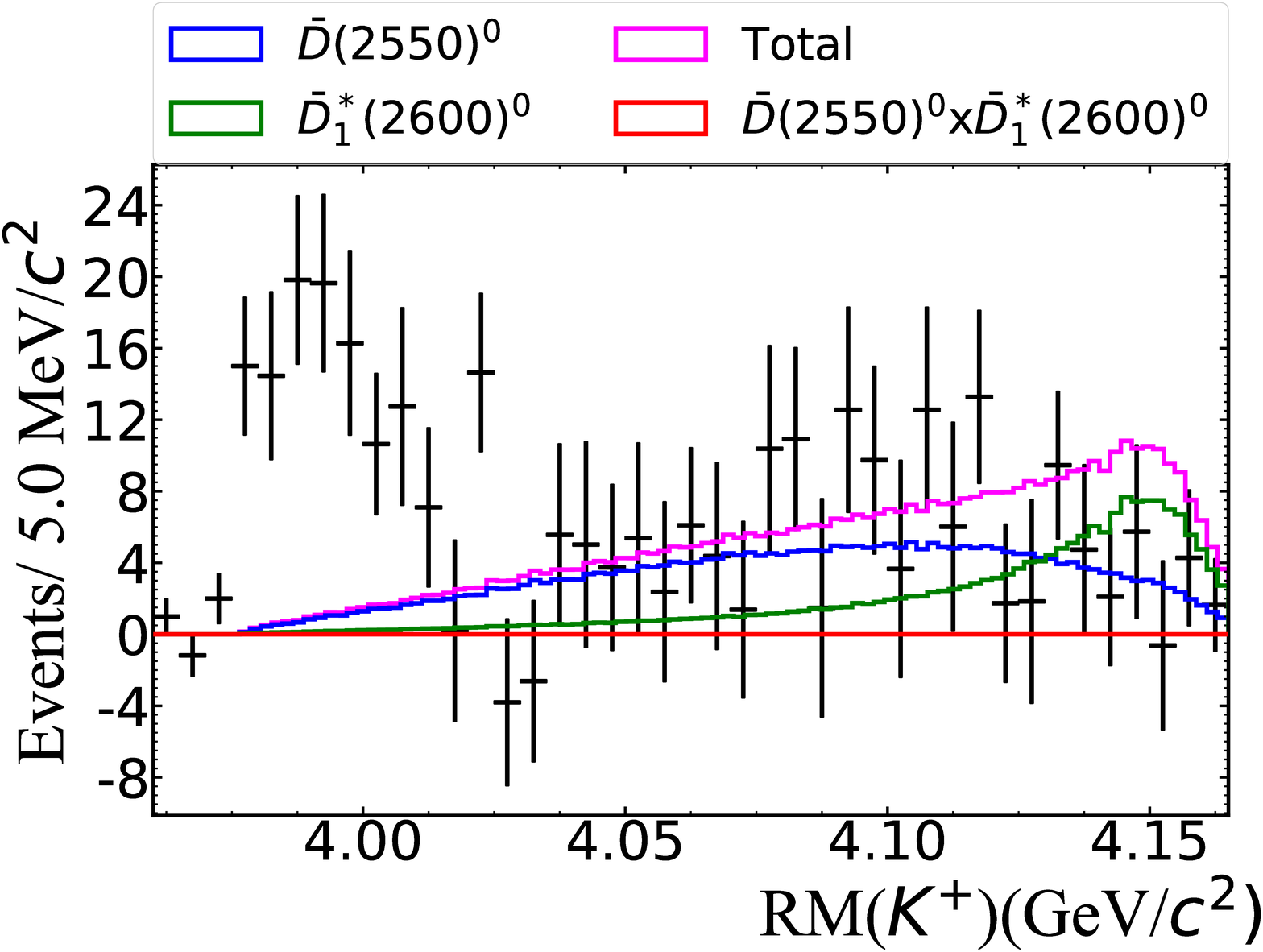}}
\subfigure[\footnotesize $\bar{D}(2550)^{0}\dzero$ and NR $1^+(S, S)$ ]
{\includegraphics[width=0.3\textwidth]{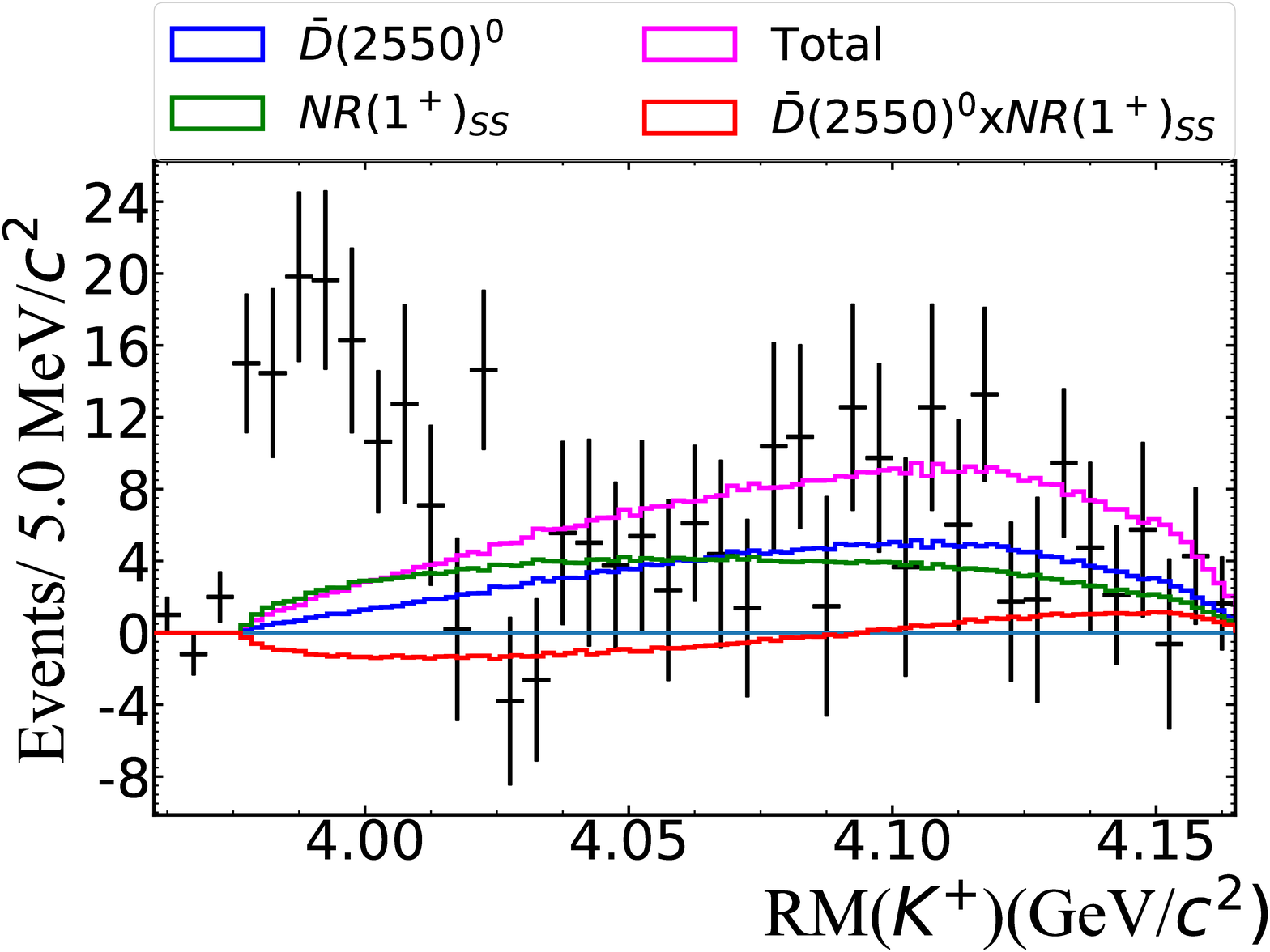}}
\subfigure[\footnotesize $\bar{D}(2550)^{0}\dzero$ and NR $1^+(D, S)$ ]
{\includegraphics[width=0.3\textwidth]{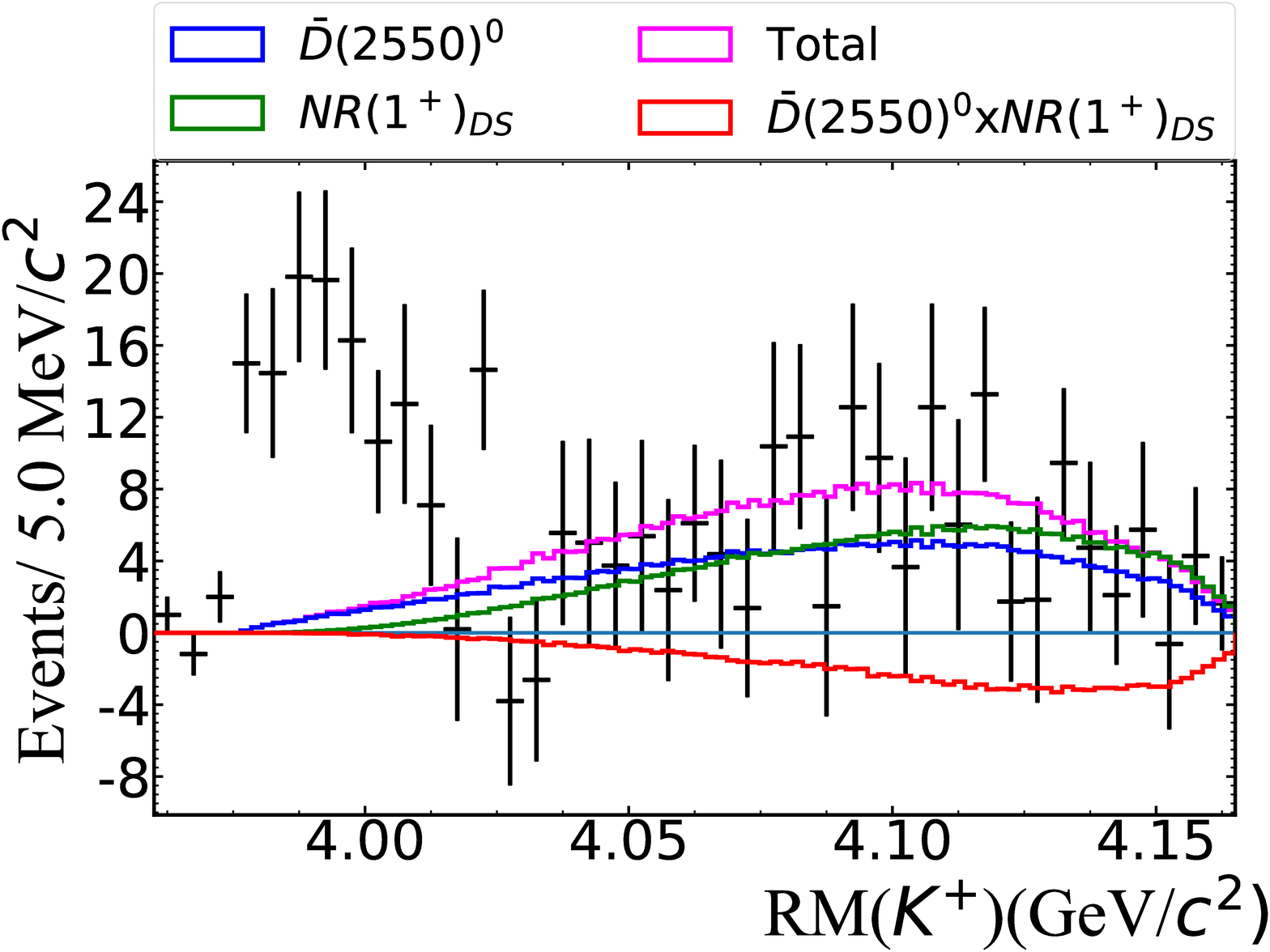}}
\subfigure[\footnotesize $D^{*}_{s2}(2573)^{+}\dsstm$ and $\bar{D}(2550)^{0}\dzero$ ]
{\includegraphics[width=0.3\textwidth]{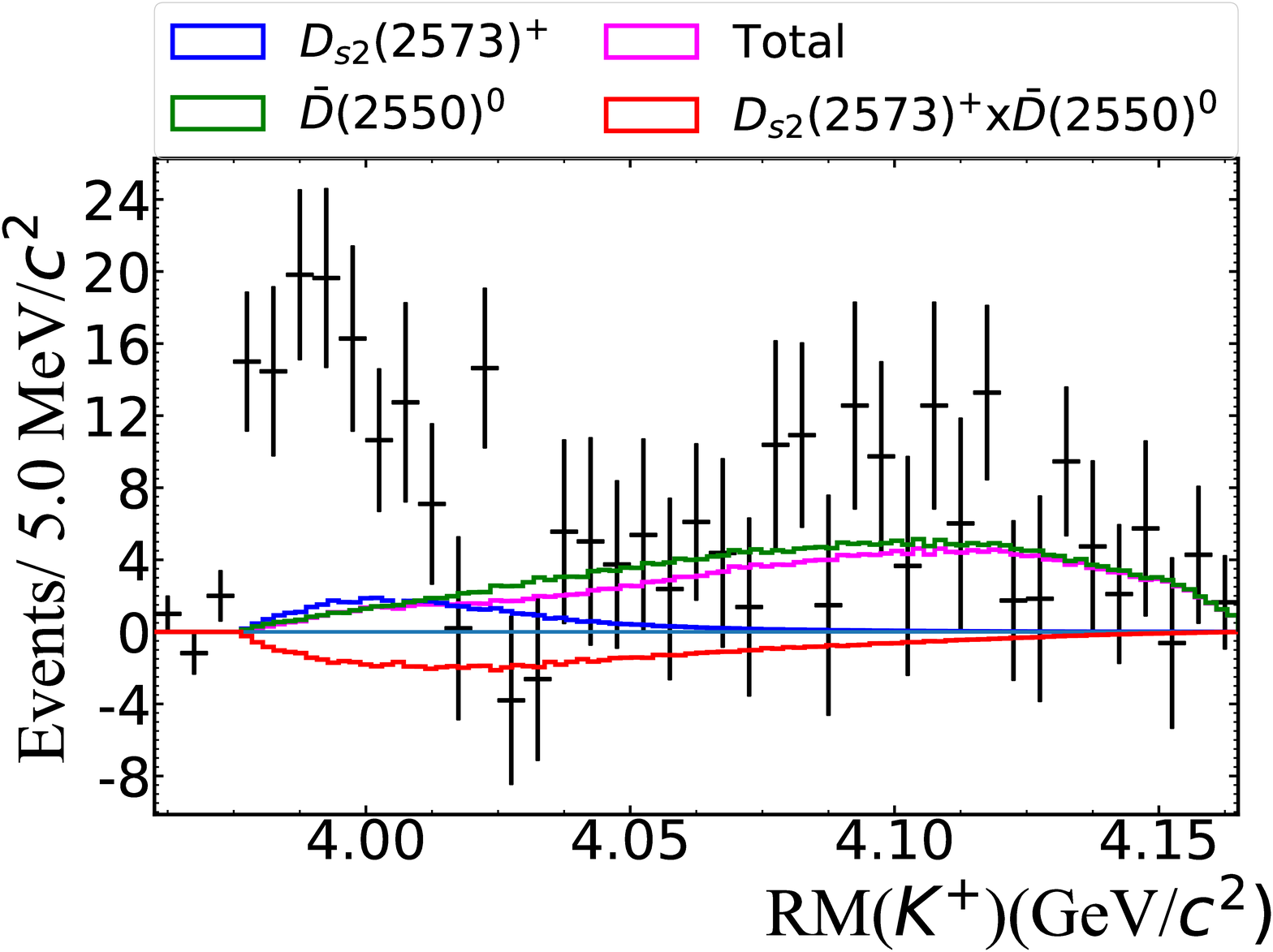}}
\subfigure[\footnotesize $D^{*}_{s2}(2573)^{+}\dsstm$ and $\bar{D}_{1}^{*}(2600)^{0}\dzero$  ]
{\includegraphics[width=0.3\textwidth]{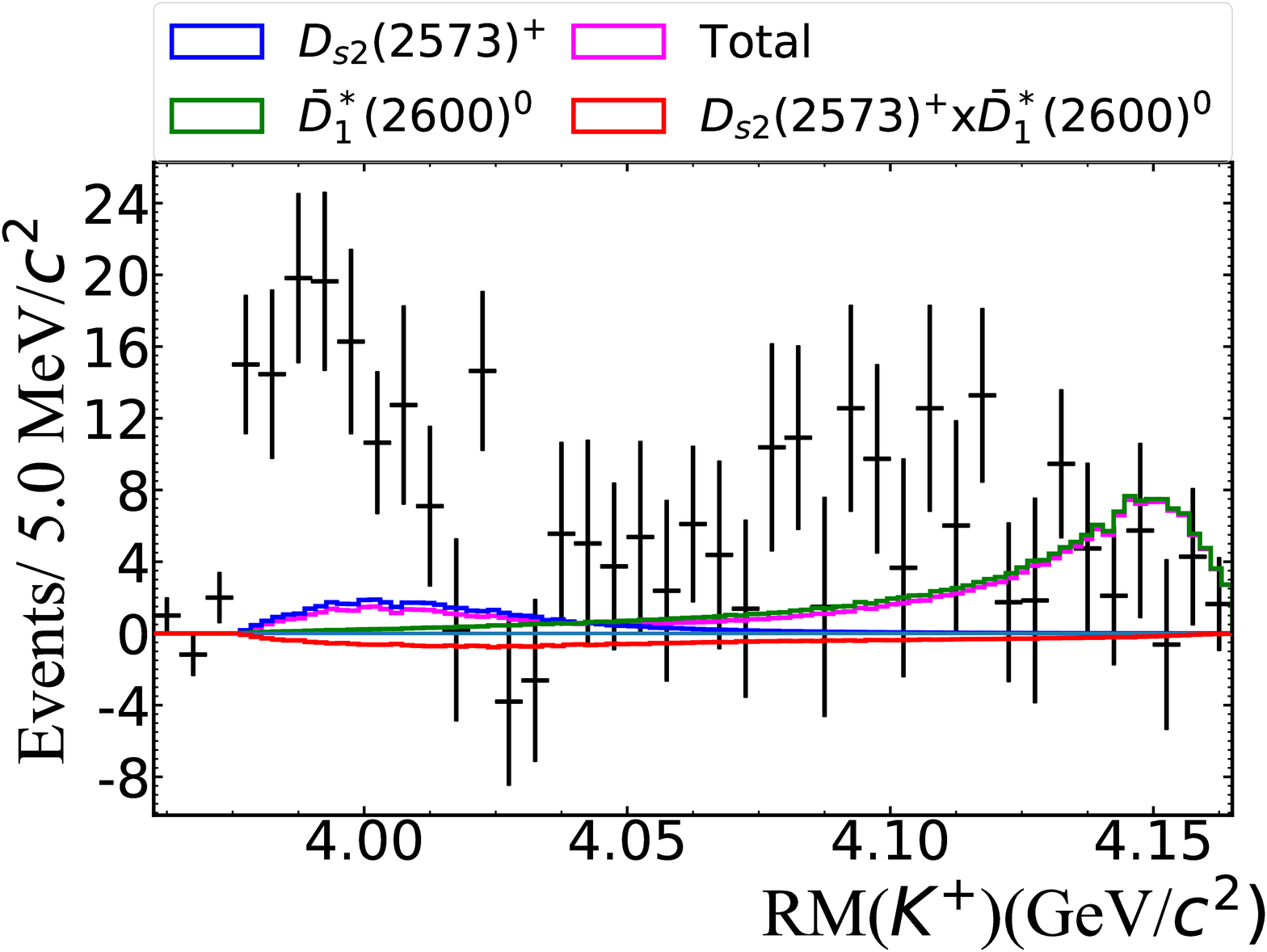}}
\subfigure[\footnotesize $D^{*}_{s2}(2573)^{+}\dsstm$ and NR $1^+(S, S)$ ]
{\includegraphics[width=0.3\textwidth]{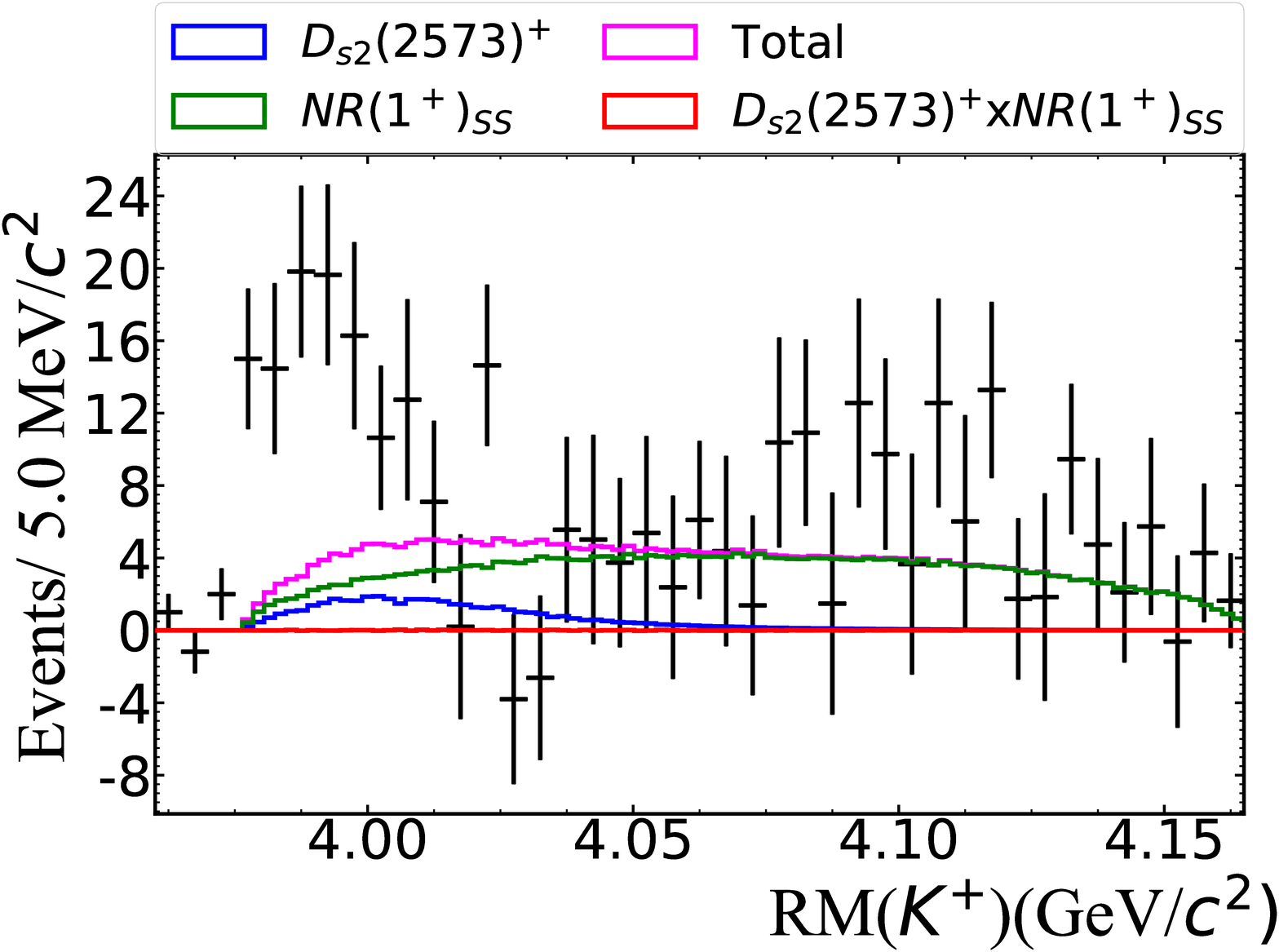}}
\subfigure[\footnotesize $D^{*}_{s2}(2573)^{+}\dsstm$ and NR $1^+(D, S)$ ]
{\includegraphics[width=0.3\textwidth]{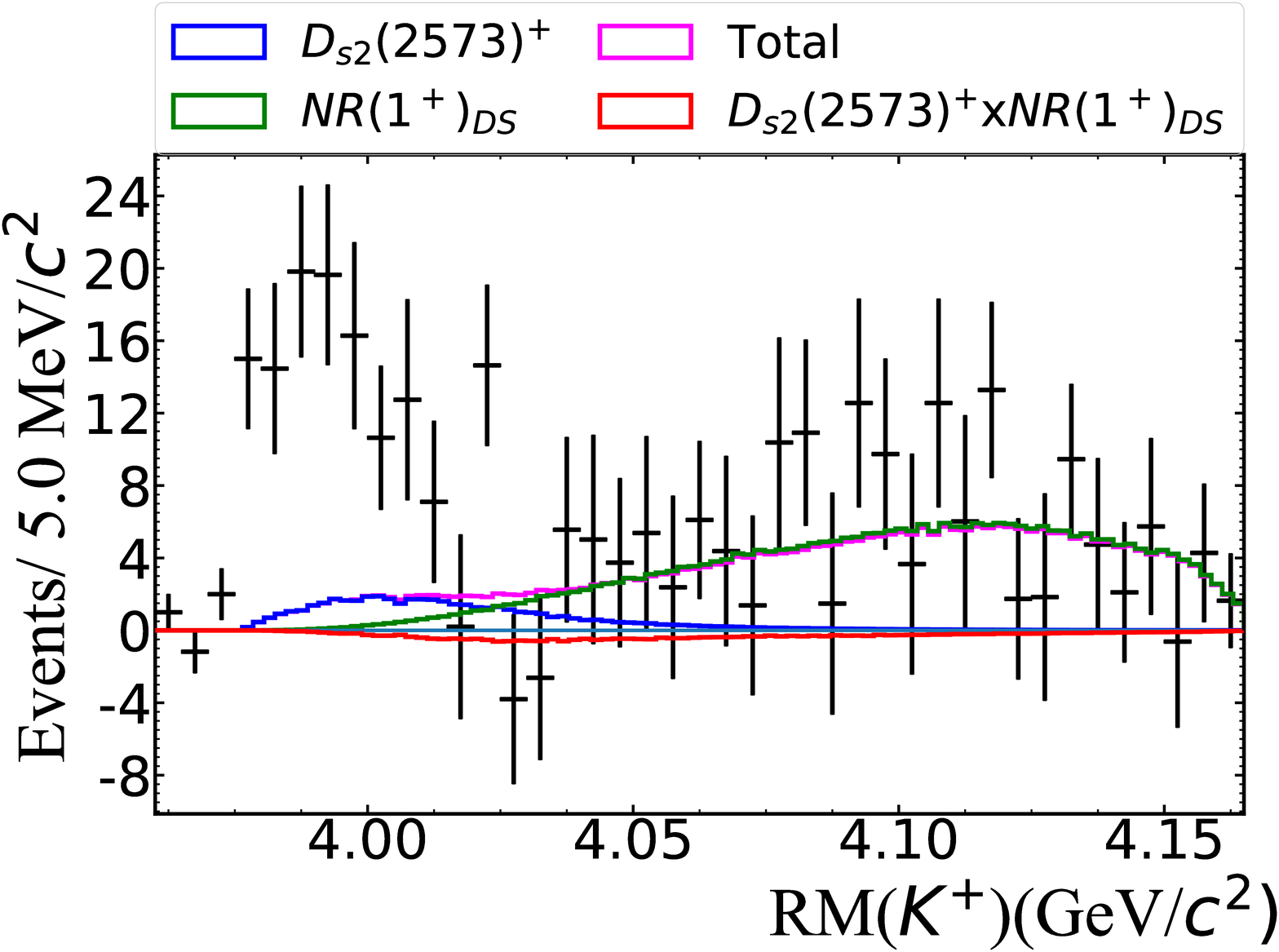}}
\subfigure[\footnotesize $\bar{D}_{1}^{*}(2600)^{0}\dzero$ and NR $1^+(S, S)$ ]
{\includegraphics[width=0.3\textwidth]{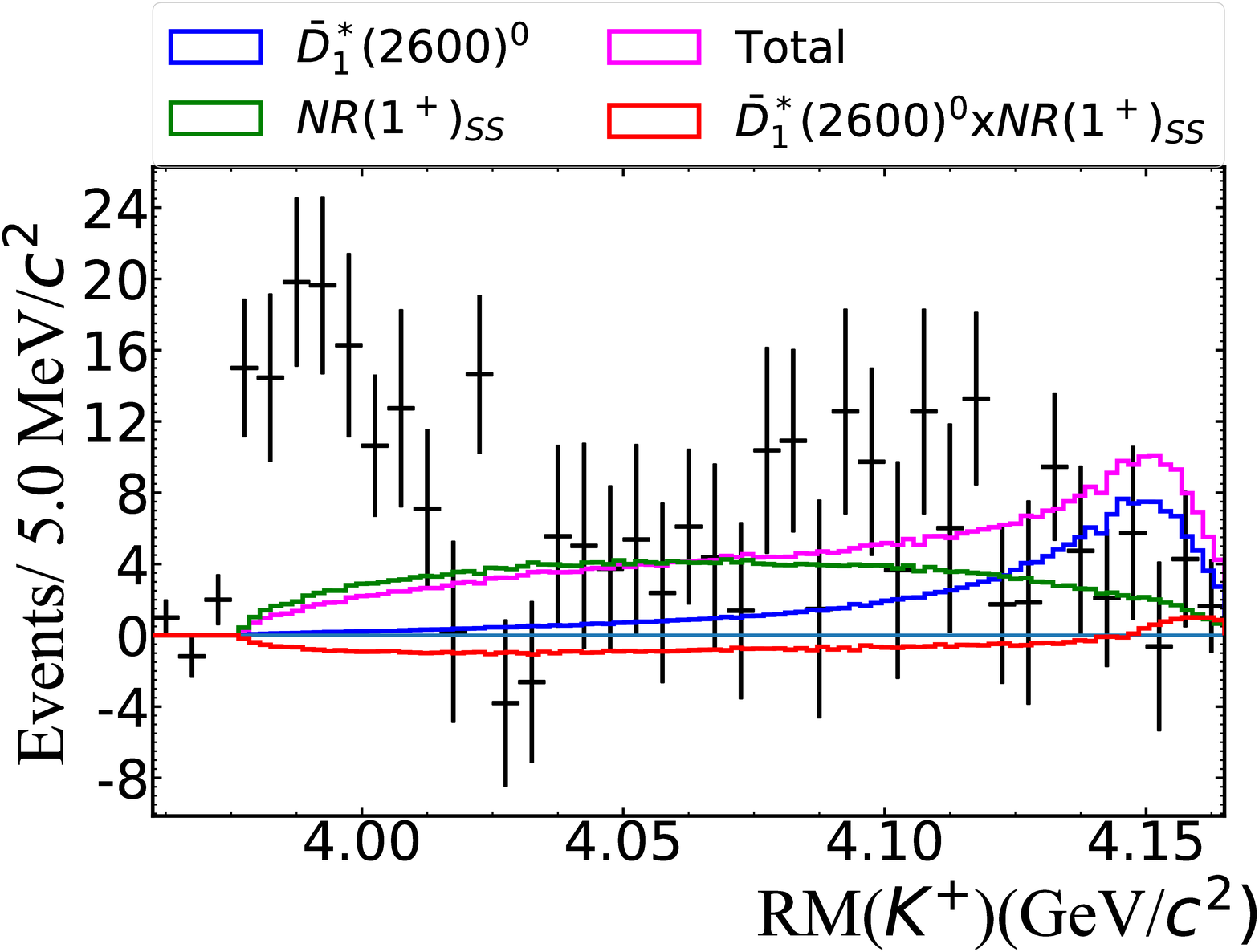}}
\subfigure[\footnotesize $\bar{D}_{1}^{*}(2600)^{0}\dzero$ and NR $1^+(D, S)$]
{\includegraphics[width=0.3\textwidth]{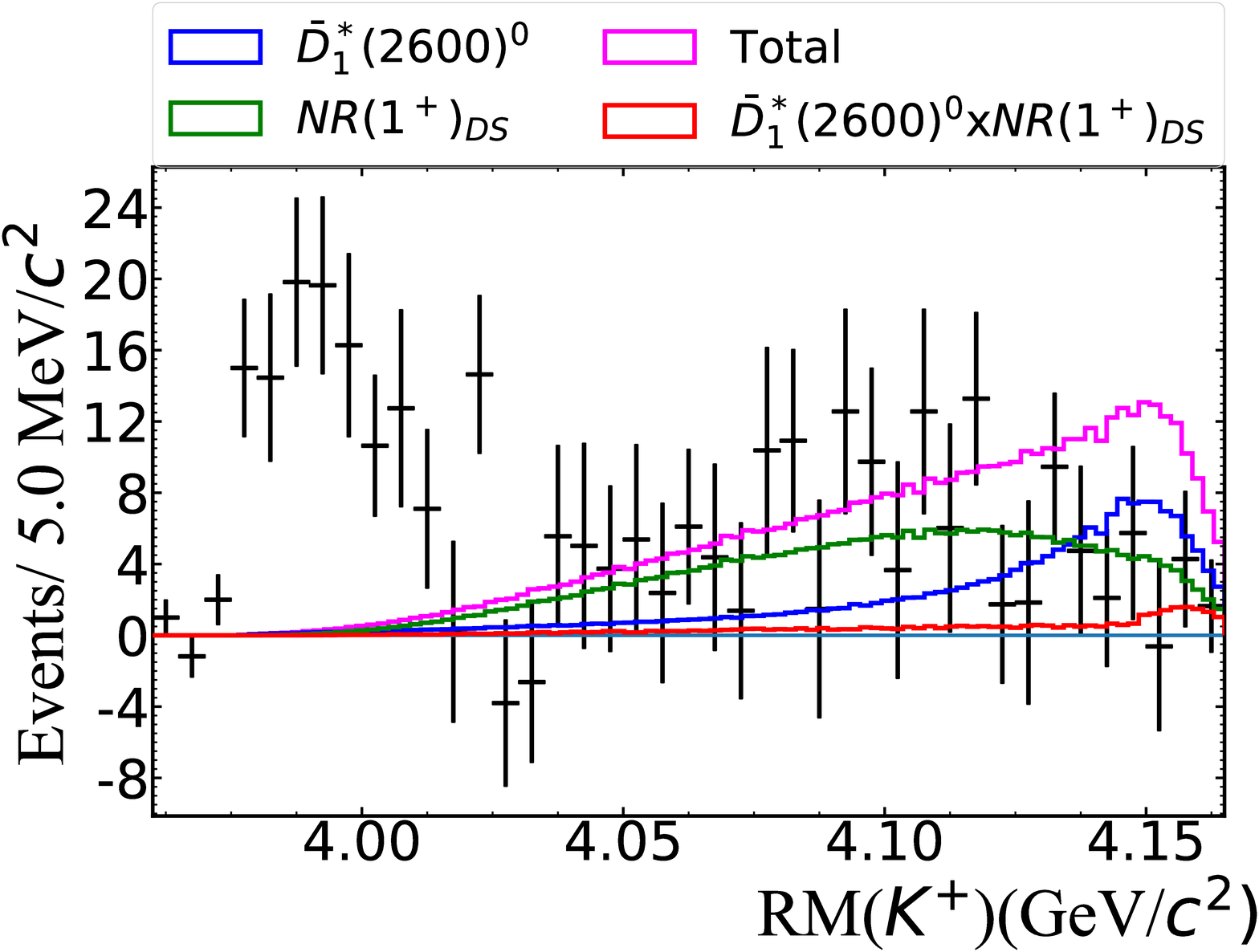}}
\subfigure[\footnotesize NR $1^+(S, S)$ and NR $1^+(D, S)$ ]
{\includegraphics[width=0.3\textwidth]{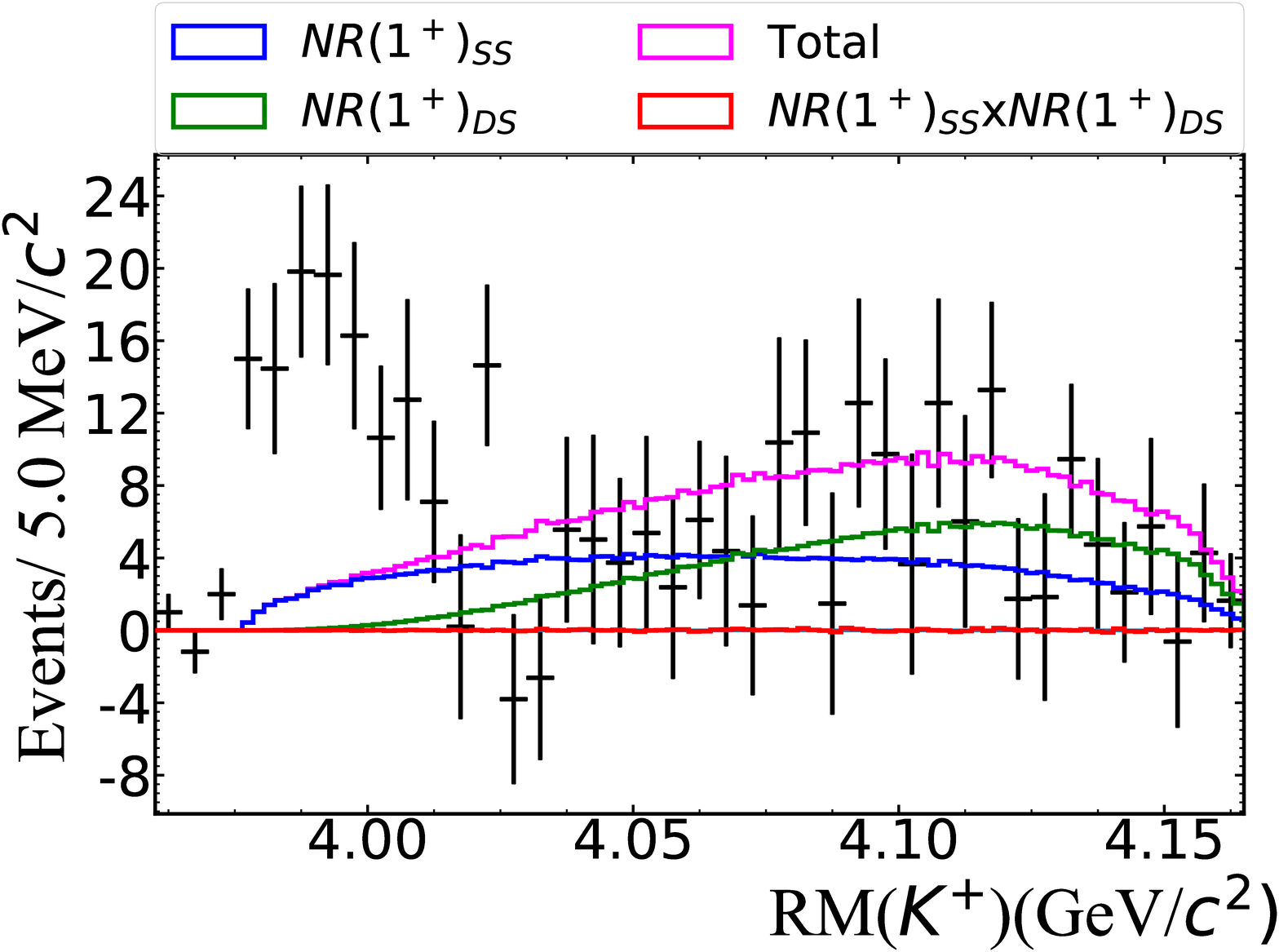}}
\caption{$\kaonp$ recoil-mass spectra in data with the WS background contributions subtracted, and MC simulations of two possible background processes for the $K^+\dsstm \dzero$  final state, whose interferences are taken into account. The interference effect is tuned to be largest around $4.0\gevcc$. In the non-resonant (NR) process, the angular momentum $(L_{\kaonp X}, L_{\dsstm \dzero})$ denotes the angular momentum between $\kaonp$ and $X_{\dsstm\dzero}$, and $\dsstm$ and $\dzero$ in the $\ee$($X_{\dsstm \dzero}$) rest frame, respectively.
Individual contributions are scaled according to the observed yields in the control samples.}
\label{fig:inter1}
\end{figure*}

\begin{figure*}[hpt]
\centering
\subfigure[\footnotesize $D_{s1}(2536)^{+}\dsm$ and $\bar{D}_{1}^{*}(2600)^{0}\dstzero$ ]
{\includegraphics[width=0.3\textwidth]{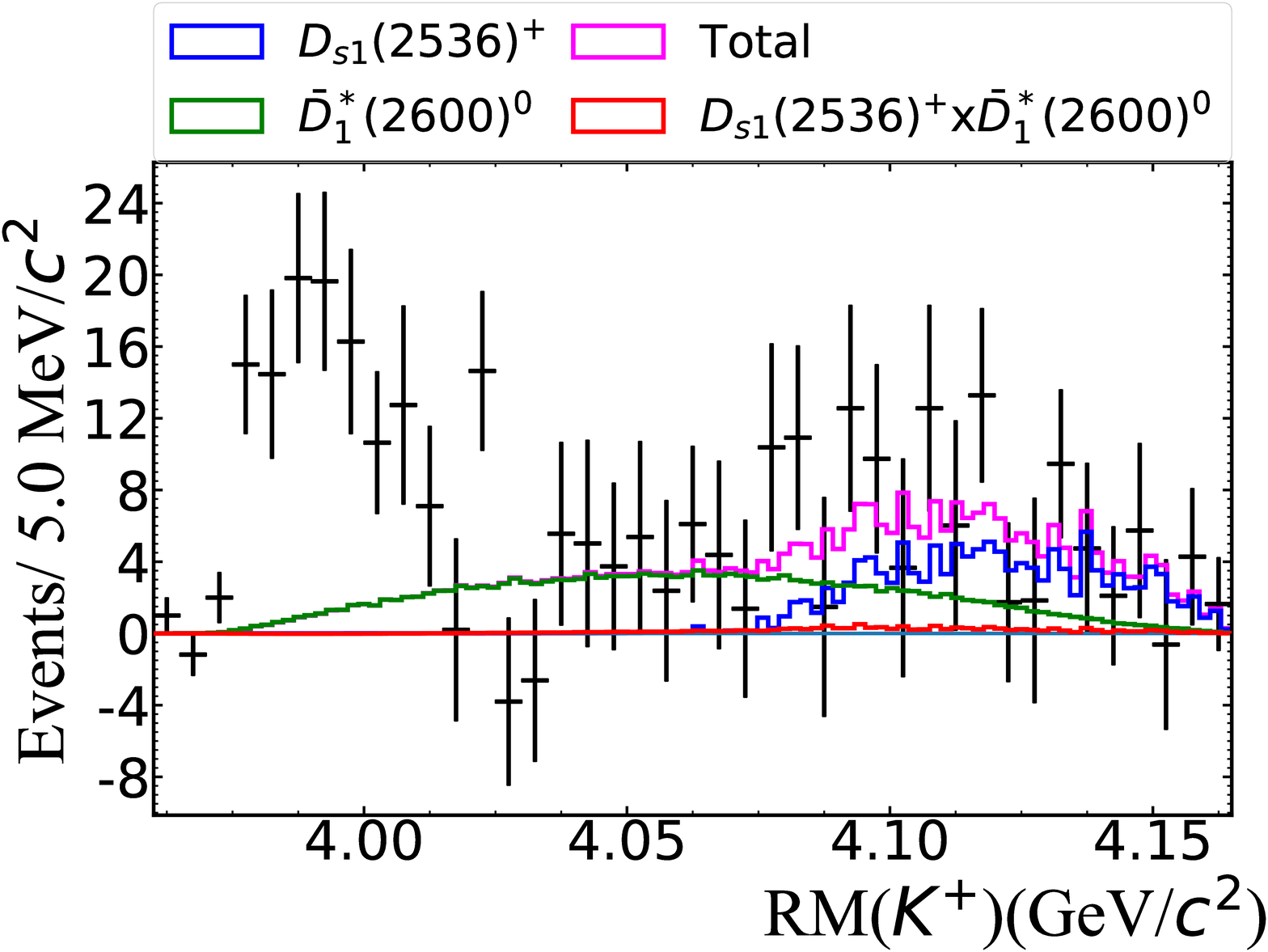}}
\subfigure[\footnotesize $D_{s1}(2536)^{+}\dsm$ and NR $1^+(S, S)$ ]
{\includegraphics[width=0.3\textwidth]{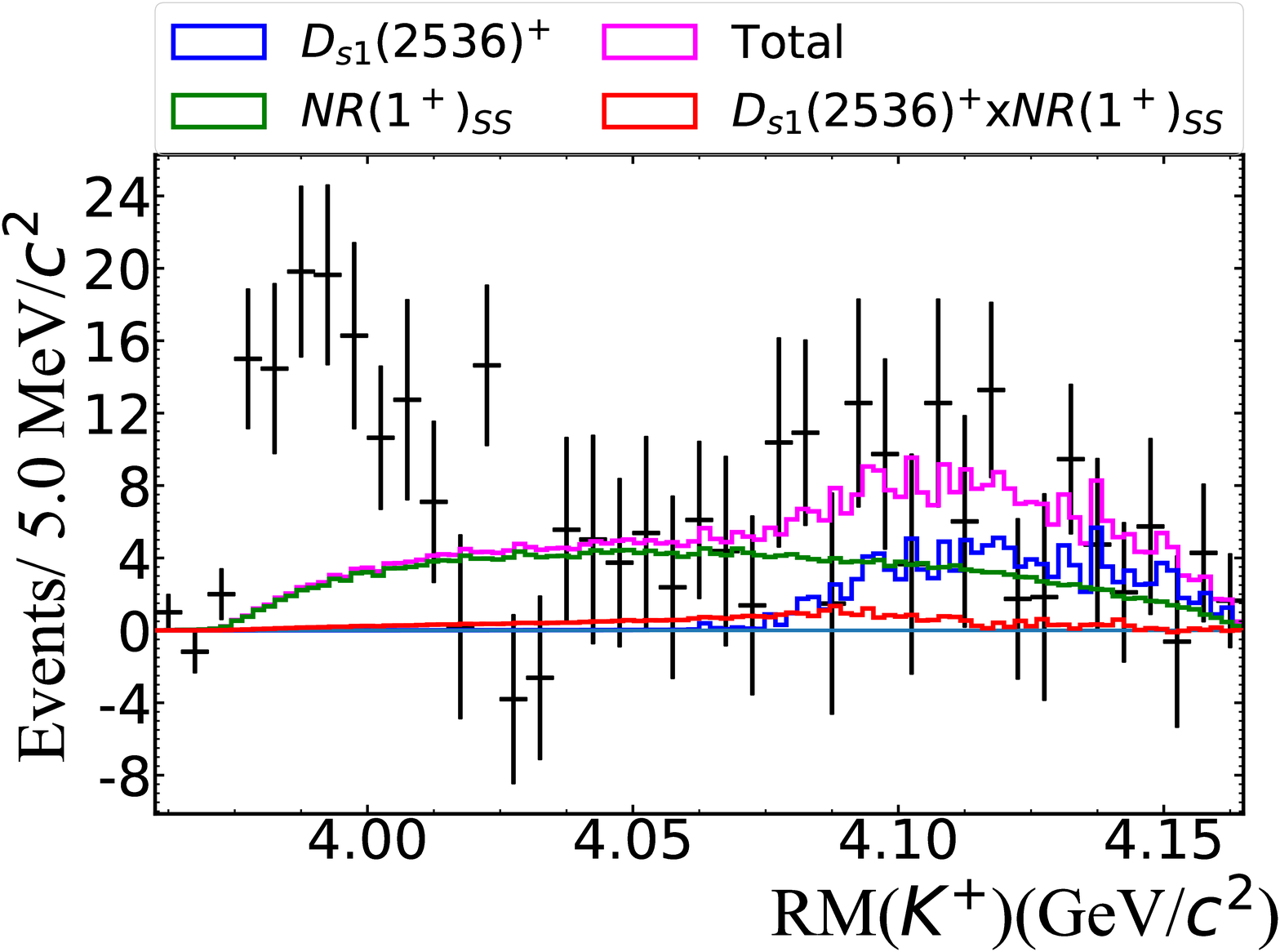}}
\subfigure[\footnotesize $D_{s1}(2536)^{+}\dsm$ and NR $1^+(D, S)$ ]
{\includegraphics[width=0.3\textwidth]{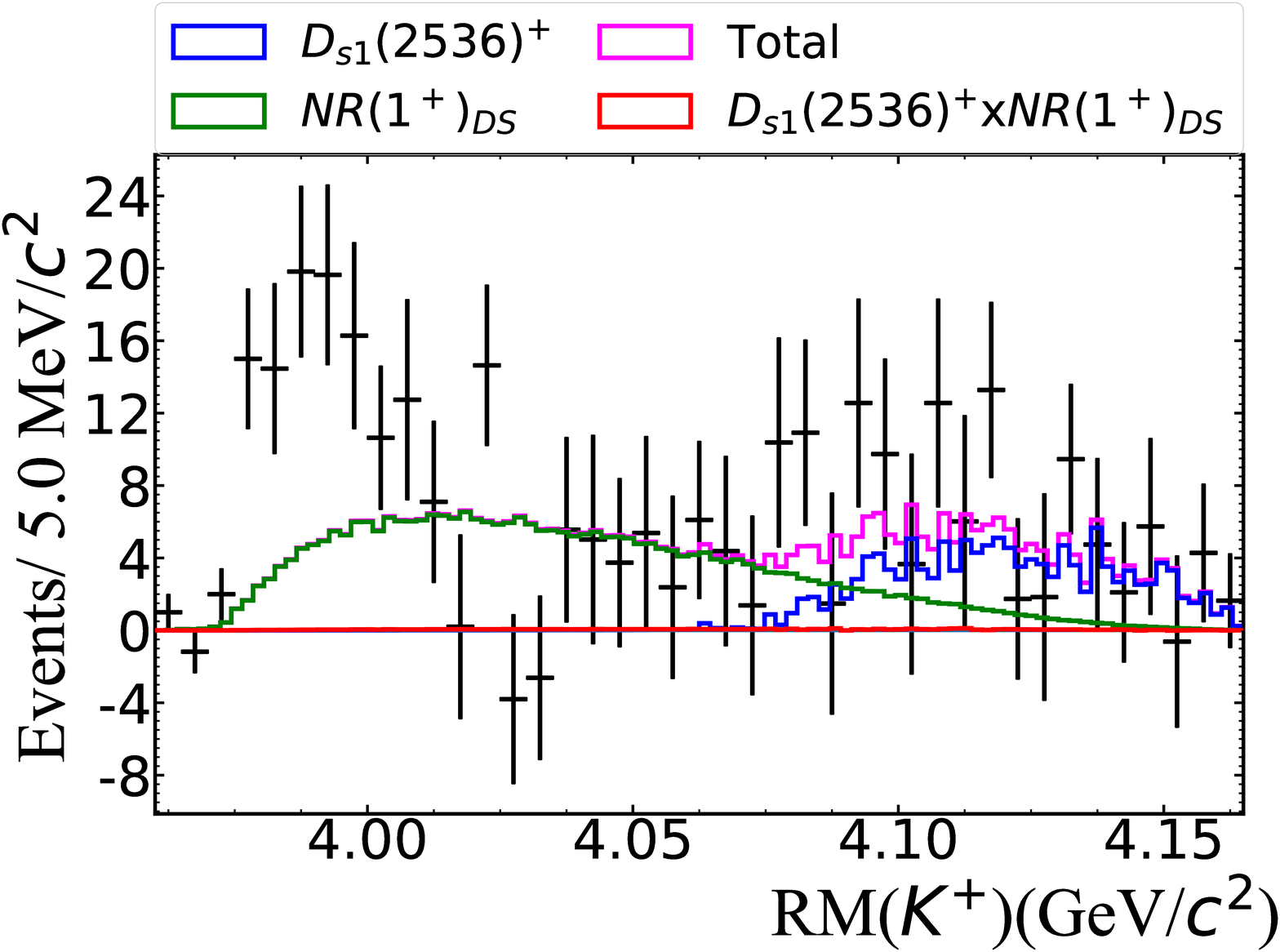}}
\subfigure[\footnotesize $\bar{D}_{1}^{*}(2600)^{0}\dstzero$ and NR $1^+(S, S)$ ]
{\includegraphics[width=0.3\textwidth]{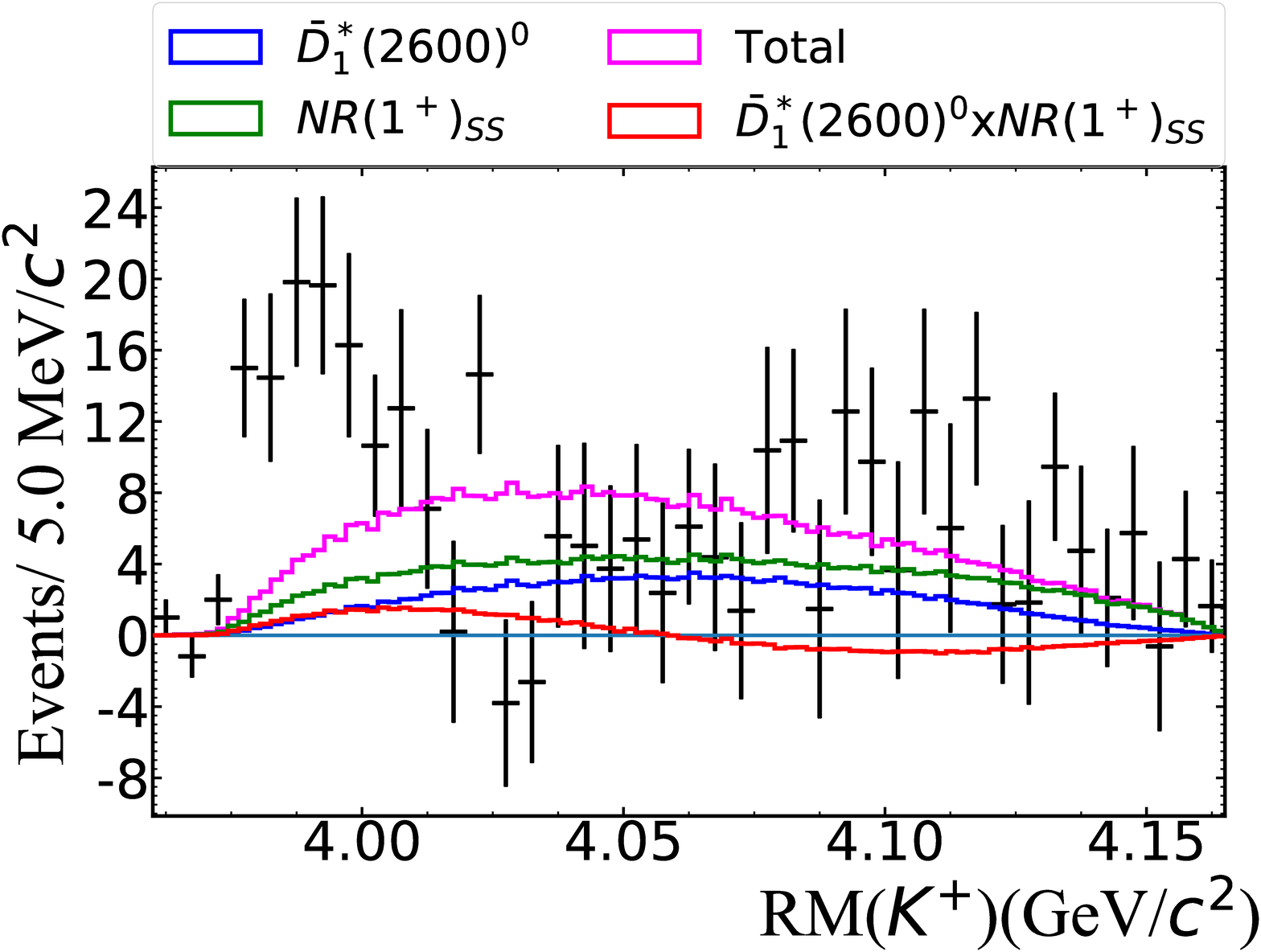}}
\subfigure[\footnotesize $\bar{D}_{1}^{*}(2600)^{0}\dstzero$ and NR $1^+(D, S)$ ]
{\includegraphics[width=0.3\textwidth]{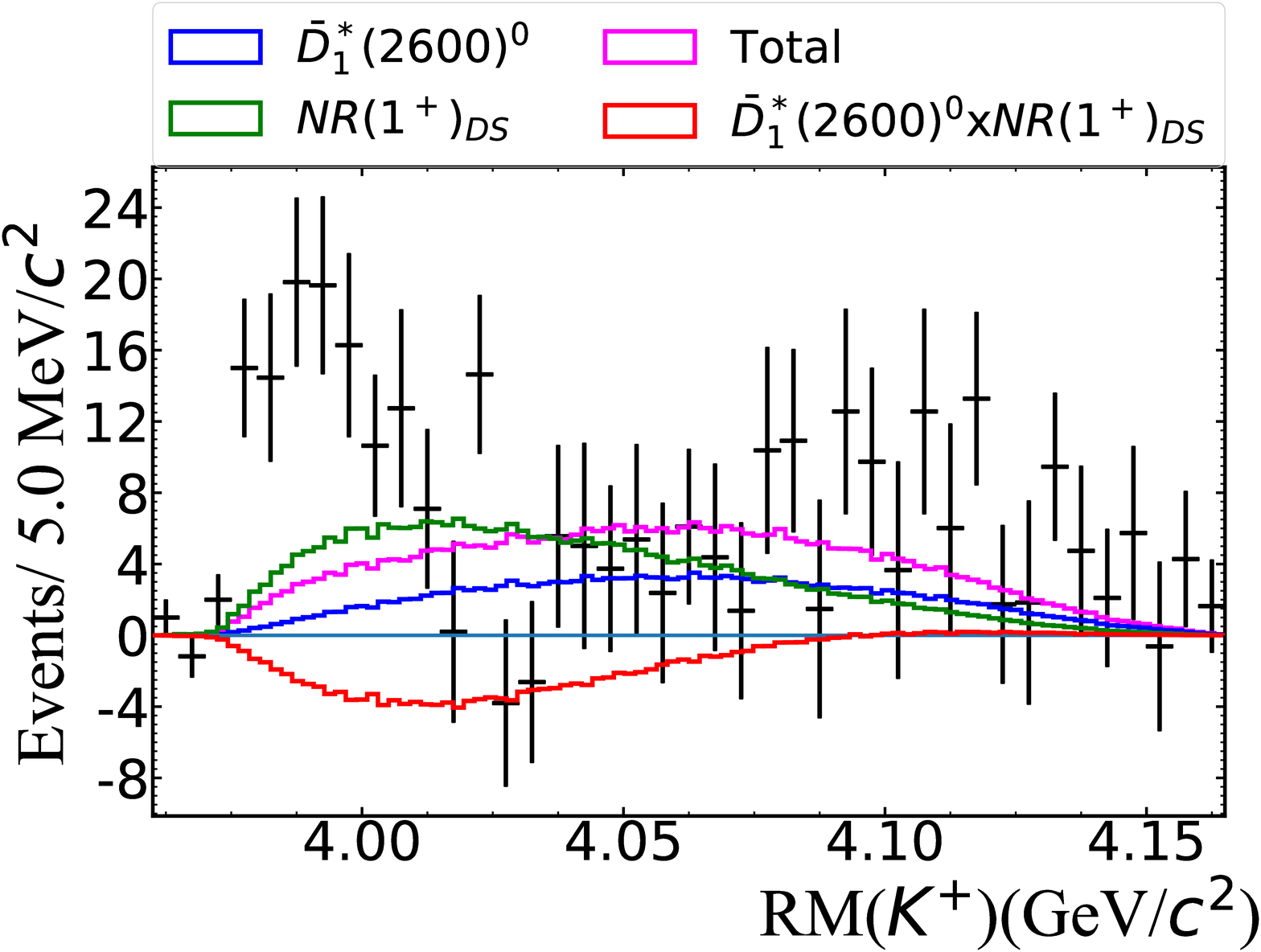}}
\subfigure[\footnotesize $D^{*}_{s1}(2700)^{+}\dsm$ and $ D_{s1}(2536)^{+}\dsm$ ]
{\includegraphics[width=0.3\textwidth]{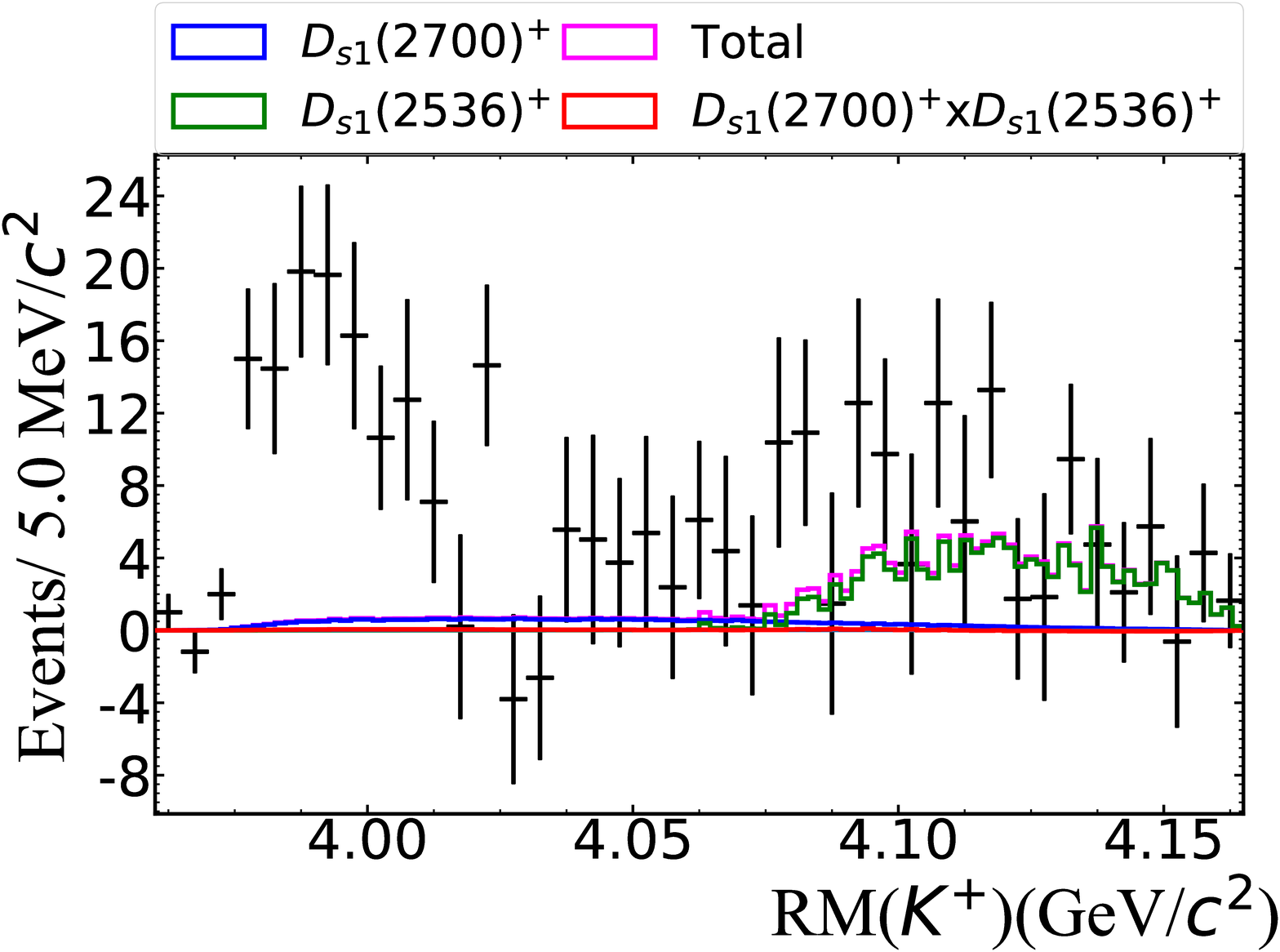}}
\subfigure[\footnotesize $D^{*}_{s1}(2700)^{+}\dsm$ and $\bar{D}_{1}^{*}(2600)^{0}\dstzero$ ]
{\includegraphics[width=0.3\textwidth]{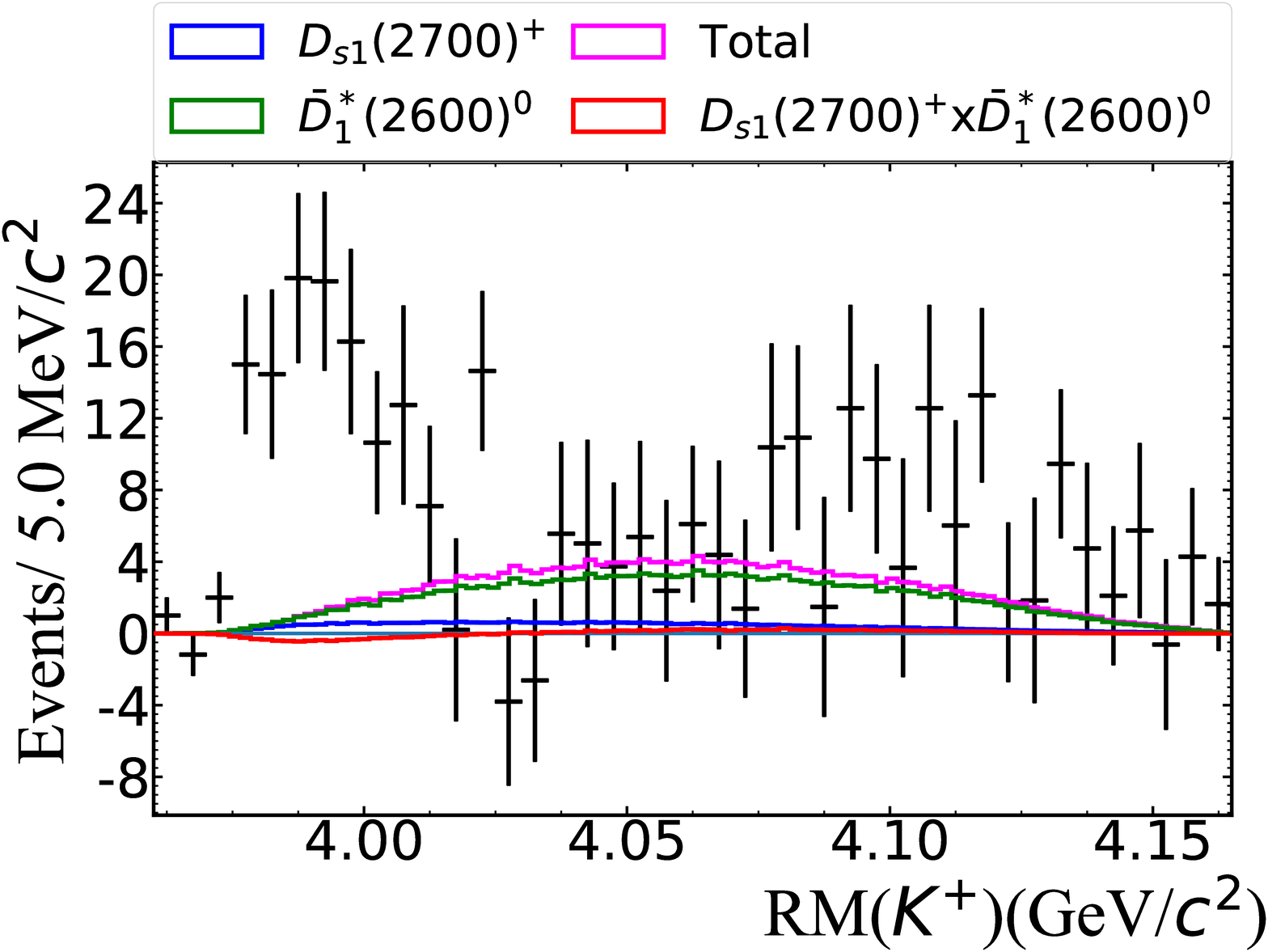}}
\subfigure[\footnotesize $D^{*}_{s1}(2700)^{+}\dsm$ and NR $1^+(S, S)$ ]
{\includegraphics[width=0.3\textwidth]{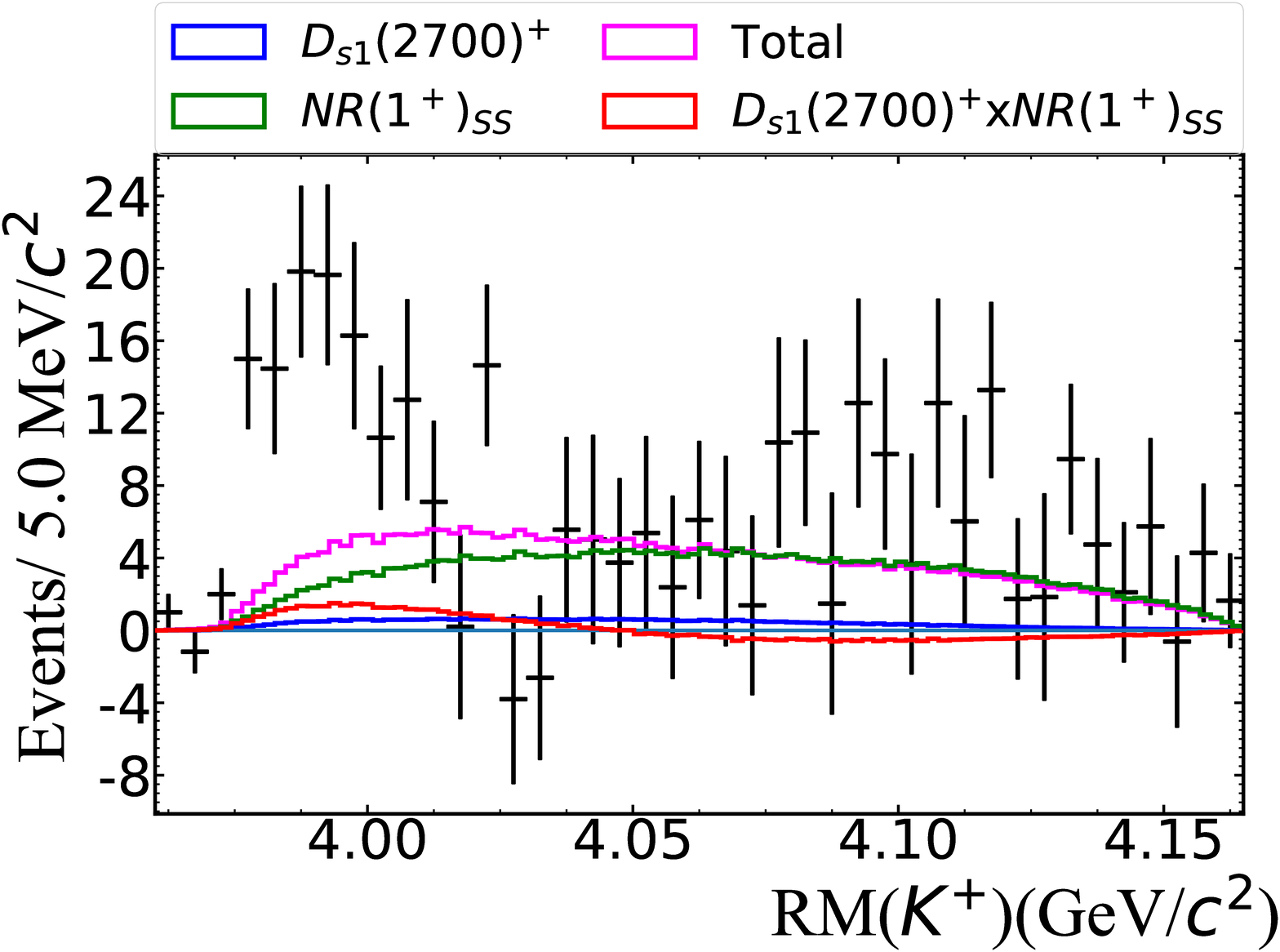}}
\subfigure[\footnotesize $D^{*}_{s1}(2700)^{+}\dsm$ and NR $1^+(D, S)$ ]
{\includegraphics[width=0.3\textwidth]{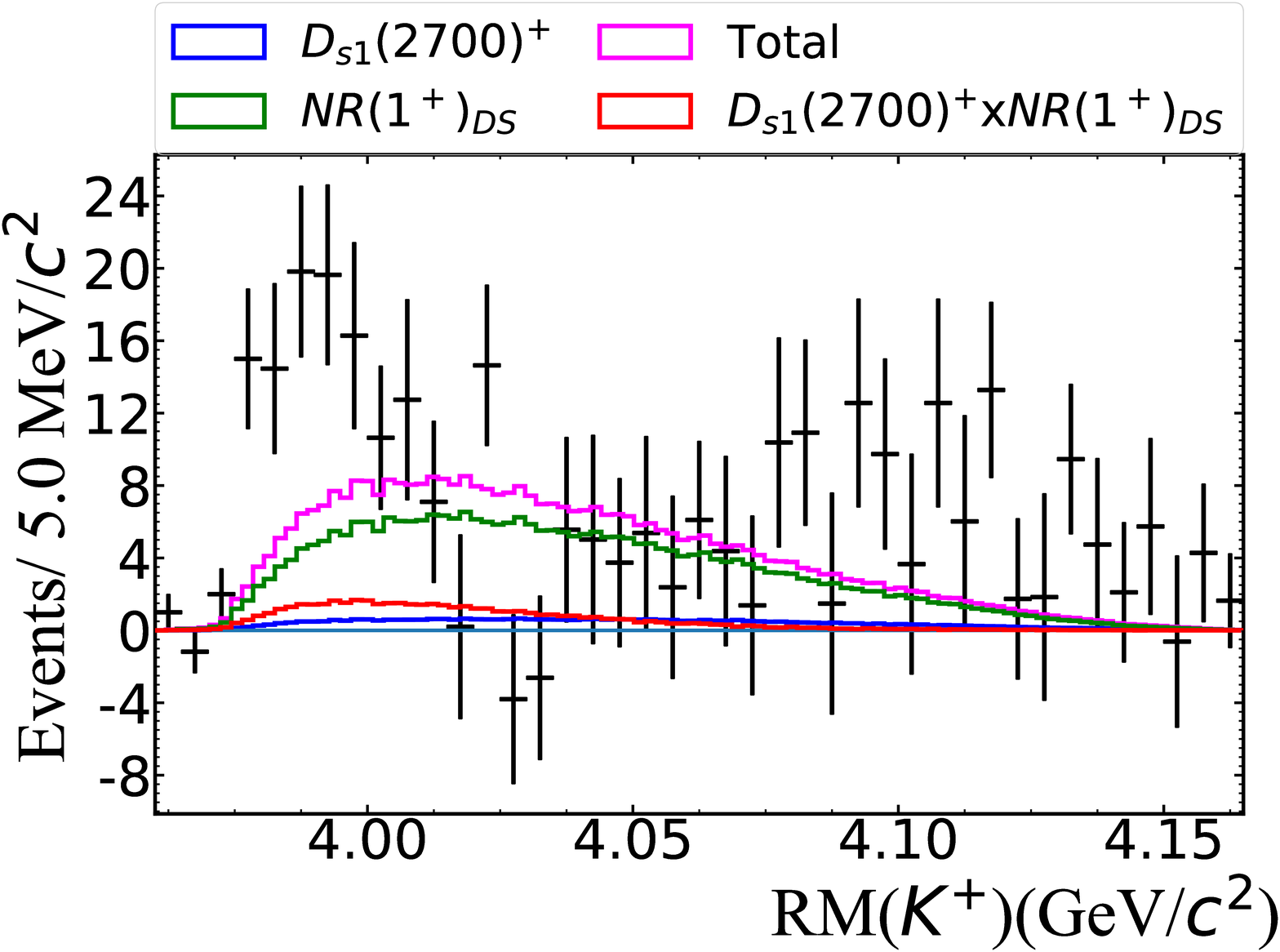}}
\subfigure[\footnotesize $D^{*}_{s1}(2700)^{+}\dsm$ and NR $1^+(D, S)$ ]
{\includegraphics[width=0.3\textwidth]{Interference/fig_Ds/fig_2700_NRDS.eps}}
\subfigure[\footnotesize NR $1^+(S, S)$ and NR $1^+(D, S)$ ]
{\includegraphics[width=0.3\textwidth]{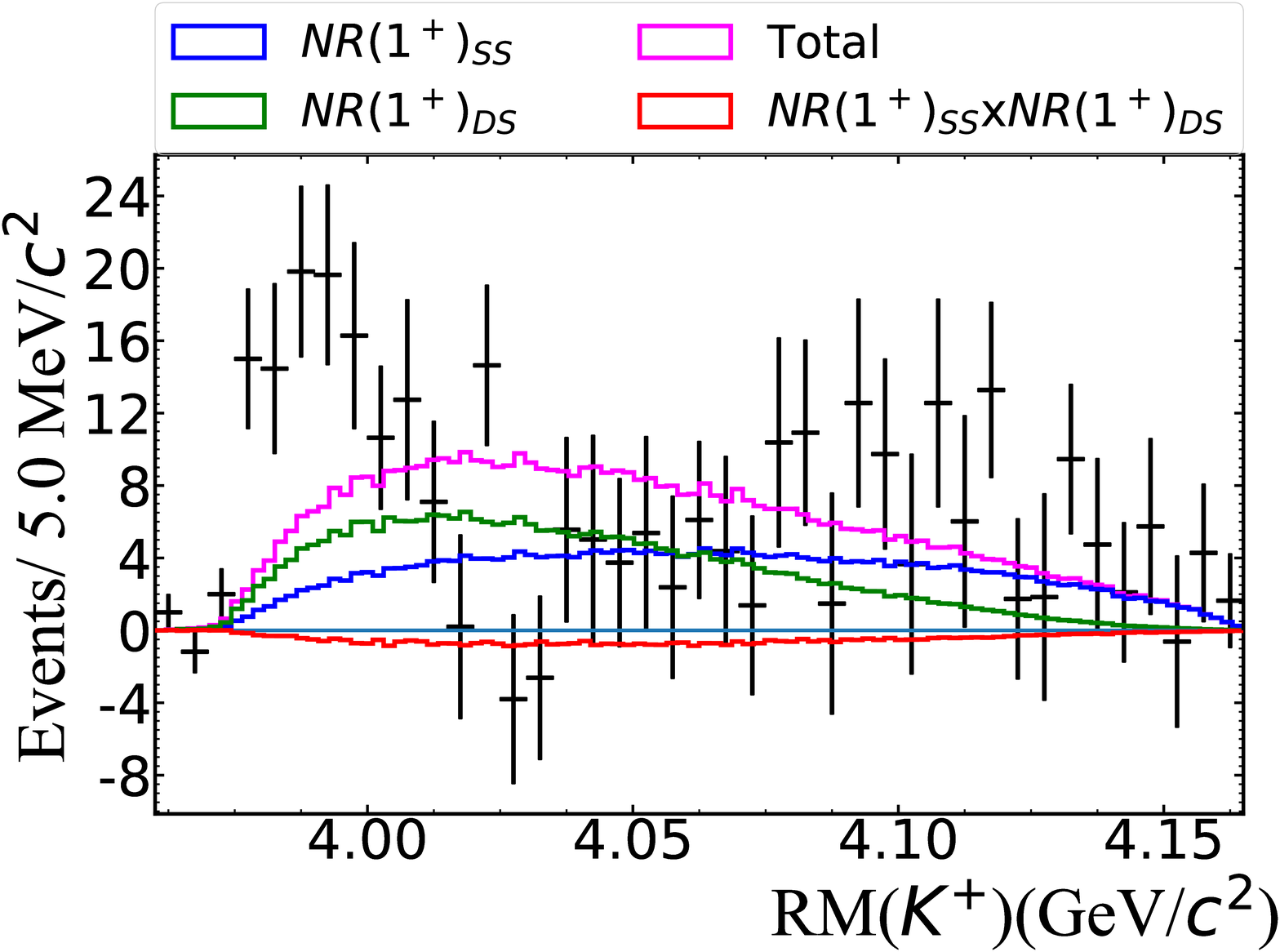}}
\caption{$\kaonp$ recoil-mass spectra in data with the WS background contributions subtracted, and MC simulations of two possible background processes for the $K^+\dsm \dstzero$  final state, whose interferences are taken into account. The interference effect is tuned to be largest around $4.0\gevcc$. In the non-resonant (NR) process, the angular momentum $(L_{\kaonp X}, L_{\dsm\dstzero})$ denotes the angular momentum between $\kaonp$ and $X_{\dsm\dstzero}$, and $\dsm$ and $\dstzero$ in the $\ee$($X_{\dsm\dstzero}$) rest frame, respectively.
Individual contributions are scaled according to the observed yields in the control samples.}
\label{fig:inter2}
\end{figure*}

To understand the potential backgrounds from excited $D_{(s)}^{**}$ states, all reported states in the PDG~\cite{pdg} whose production and decay is allowed within the available phase-space at $\sqrt{s}=4.681$ GeV are investigated.
The corresponding $RM(\kaonp)$ distributions of the MC simulations are plotted in Figs.~\ref{fig:DDh_RMK_Ds} and \ref{fig:DDh_RMK_D}.
Furthermore, possible interferences among those excited  $D_{(s)}^{**}$ states are systematically scanned, and the choices with the largest interferences around $RM(\kaonp)=4.0 \gevcc$ are compared with the distributions in data, shown in Fig.~\ref{fig:inter1} and Fig.~\ref{fig:inter2}. It is evident that none of the states can explain the narrow peaking structure below $4.0 \gevcc$.

\clearpage
\section{Systematics uncertainties}

Sources of systematic uncertainties on the measurement of the $\zcsm$ resonance parameters and the cross section are studied, in which the main sources include the mass scaling, detector resolution, the signal model, background models and the input cross-section line shape for $\sigma^{B}(\ee\to \kaonp\zcsm)$.

We select a control sample of $\ee\to D_{s1}(2536)^+ \dsstm\to\kaonp\dstzero\dsstm$ at $\sqrt{s} =4.681\gev$ by detecting $K^+D^{*0}$ with $D^{*0}\to \pi^0 D^0$, $D^0\to K^-\pi^+$, $K^-\pi^+\pi^0$ as well as $K^-\pi^+\pi^+\pi^-$ with a missing $D_s^{*-}$ in the final state to study the mass scaling of the recoil mass of the low-momentum bachelor $\kaonp$.
We fit the $\dsstm$ peak in the spectra of the recoil mass of $\kaonp\dstzero$, where the  $\dsstm$ signal is modeled with a MC-determined signal shape convolved with a Gaussian function to represent a potential difference between data and MC simulation.
The fitted Gaussian parameters are determined to be $\mu=-0.2\pm0.5\mevcc$ and $\sigma_{\rm upper}<1.43\mev$ (68\% C.L.), which are used to determine the systematic effects due to mass scaling and detection resolution. After incorporating the evaluated detection resolution difference up to the upper uncertainty, we find the maximum change on the result of the fitted width to be $1.0\mev$.

In this work the two $Z_{cs}$ signal processes are difficult to distinguish due to the partial-reconstruction method and the limited sample size. Hence, without any a priori knowledge, we vary the BF ratio $f$ in the range from 0.2 to 0.8, corresponding to the standard deviation of a uniform distribution from 0 to 1.
We find the resulting changes on the mass and width to be $0.2\mevcc$ and $1.0\mev$, respectively.
In the nominal fit, we assume that the spin-parity of the $\zcsm$ is $1^+$ and that the relative momentum between $\kaonp$ and $\zcsm$ in the rest frame of the $\ee$ system and the relative momentum between $\dsm$($\dsstm$) and $\dstzero$($\dzero$) in $\zcsm$ system are both in an $S$-wave state, denoted as $1^+$($S$, $S$).
This hypothesis can only be verified by an amplitude analysis of the signal final states, which is not feasible with the current statistics.
Therefore, as systematic variations, we test the assumptions of spin-parity and angular momentum with $1^+$($D$, $S$), $0^-$($P$, $P$), $1^-$($P$, $P$) and $2^-$($P$, $P$) configurations. These tests give maximum changes of $1.0\mevcc$ in the mass and $2.6\mev$ in the width.
The systematic uncertainty related to the combinatorial background is estimated by varying both the sideband yield within its uncertainties and the background parametrization;
the quadrature sums of each largest difference from the nominal fit are $0.5\mevcc$ and $0.5\mev$ for the mass and width, respectively, which are taken as the systematic uncertainties.
The efficiency curves adopted in the resonance fit are varied within the uncertainties of their parametrizations, and the differences of $0.1\mevcc$ in mass and $0.2\mev$ in width to the nominal fit are taken as the related systematic uncertainty.

Any potential effects of the known $D_{(s)}^{**}$ states (as listed in Table~\ref{tab:DDh_size}) on the measurements are evaluated. We vary the size of the $D^{**+}_{s}$ and $\bar{D}^{*}_{3}(2750)^{0}$ background components within their uncertainties in the fit and take the variations as systematic uncertainties.
For the known $\bar{D}^{**0}$ states, which have $RM(K^+)$ distributions similar to that of the NR signal, the fit is repeated with each state as an additional component with its shape taken from MC simulation and the yield as a free parameter.
To further check the $\bar{D}_1^{*}(2600)^0$ component, we remove the NR component from the simultaneous fit.  The ratio
$\br{\bar{D}_1^{*}(2600)^0\to D_s^- K^+}/\br{\bar{D}_1^{*}(2600)^0\to D^- \pi^+}$
then increases from $0.00\pm0.02$ to $0.12\pm0.02$.
We evaluate the quadrature sum of the mass and width differences between each of the results from these alternative fits with respect to the nominal fit and assign the quadrature sums as related systematic uncertainties of $1.0\mevcc$ for the mass and $3.4\mev$ for the width.
We vary the input Born cross section $\sigma^B(\ee\to \kaonp\zcsm)$ within the uncertainties and repeat the signal extraction, which gives a maximum change of $0.6\mevcc$ for the mass and $1.7\mev$ for the width.

Other systematic effects mostly influence the measurement of the cross section.
Average uncertainties associated with the tracking, PID and $\kshort$ reconstruction efficiencies are estimated to be 3.6\%, 3.6\% and 0.4\%, respectively.
The efficiency of the $RM(\kaonp\dsm)$ requirement is re-estimated by changing the MC-simulated resolution according to the observed difference with respect to data and the resulting change is taken as the systematic uncertainty on the cross section.
The integrated-luminosity uncertainty, measured with large-angle Bhabha events, is estimated to be 1\%.
The uncertainties on the quoted BFs for the involved decays~\cite{pdg} are included as part of the systematic uncertainty.

\begin{table}[ph]
  \begin{center}
   \caption{Summary of systematic uncertainties on the cross sections at different energy points. The total systematic uncertainty corresponds to a quadrature sum of all individual items.}
  \begin{tabular}{l|ccccc}
      \hline \hline
      Source   &  $\sigma_{4.628}\mathcal{B}$(\%) &  $\sigma_{4.641}\mathcal{B}$(\%) &  $\sigma_{4.661}\mathcal{B}$(\%) &  $\sigma_{4.681}\mathcal{B}$(\%) &  $\sigma_{4.698}\mathcal{B}$(\%) \\ \hline
      Tracking                  & 3.6 & 3.6 & 3.6 & 3.6 & 3.6 \\
      Particle ID               & 3.6 & 3.6 & 3.6 & 3.6 & 3.6 \\
      $\kshort$                 & 0.4 & 0.4 & 0.4 & 0.4 & 0.4 \\
      $RM(\kaonp\dsm)$          & 4.0 & 0.3 & 0.4 & 0.6 & 0.2 \\
      Resolution                & 0.2  & 1.0  & 1.9  & 1.1  & 0.8  \\
      $f$ factor                & 7.8  & 7.7  & 6.7  & 6.4  & 5.9  \\
      Signal model              & 20.5 & 14.4 & 16.6 & 21.9 & 11.2 \\
      Backgrounds               & 54.8 & 5.9  & 12.0 & 3.1  & 7.8  \\
      Efficiencies              & 0.2  & 0.2  & 0.2  & 0.5  & 0.1  \\
      $D_{(s)}^{**}$ states     & 47.1 & 82.2 & 35.3 & 15.7 & 35.3 \\
      $\sigma^B(\kaonp\zcsm)$   & 11.9 & 5.7  & 22.1 & 13.4 & 32.1 \\
      Luminosity                & 1.0  & 1.0  & 1.0  & 1.0  & 1.0  \\
      Input BFs                 & 2.7  & 2.7  & 2.7  & 2.7  & 2.7  \\
          \hline
          total                 & 76.8 & 84.5 & 47.3 & 31.5 & 50.3 \\
        \hline\hline
    \end{tabular}
    \label{tab:syst2}
  \end{center}
  \end{table}

Table~\ref{tab:syst2} summarizes the systematic uncertainties on the cross sections at $\sqrt{s}$=4.628, 4.641, 4.661, 4.681 and 4.698$\gev$.

\clearpage

\section{Fit results based on three subsets of data set at $\sqrt{s}=\;$4.681~${\rm\bf GeV}$}

\begin{figure}[h]
\centering
\includegraphics[width=0.28\linewidth]{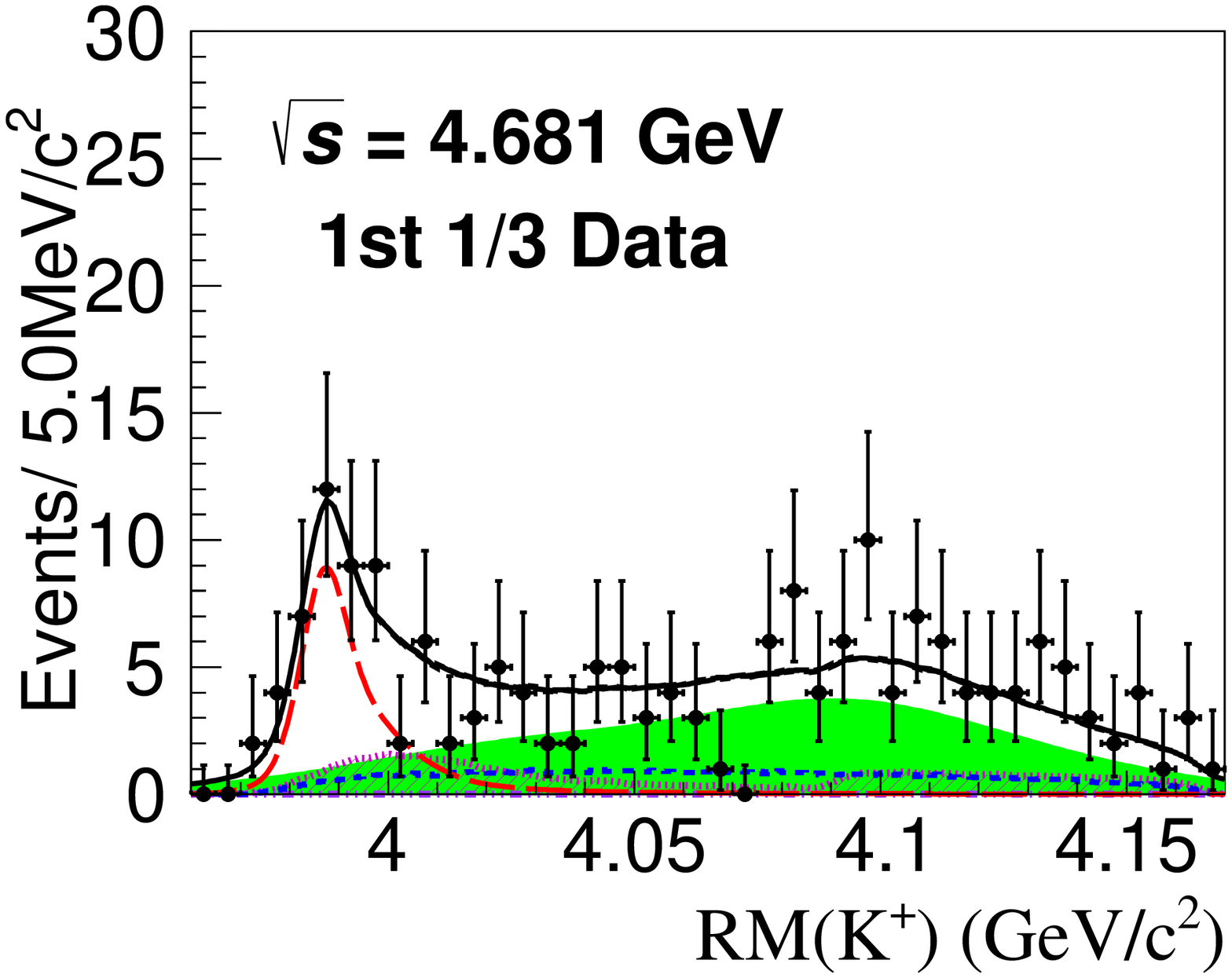}
\includegraphics[width=0.28\linewidth]{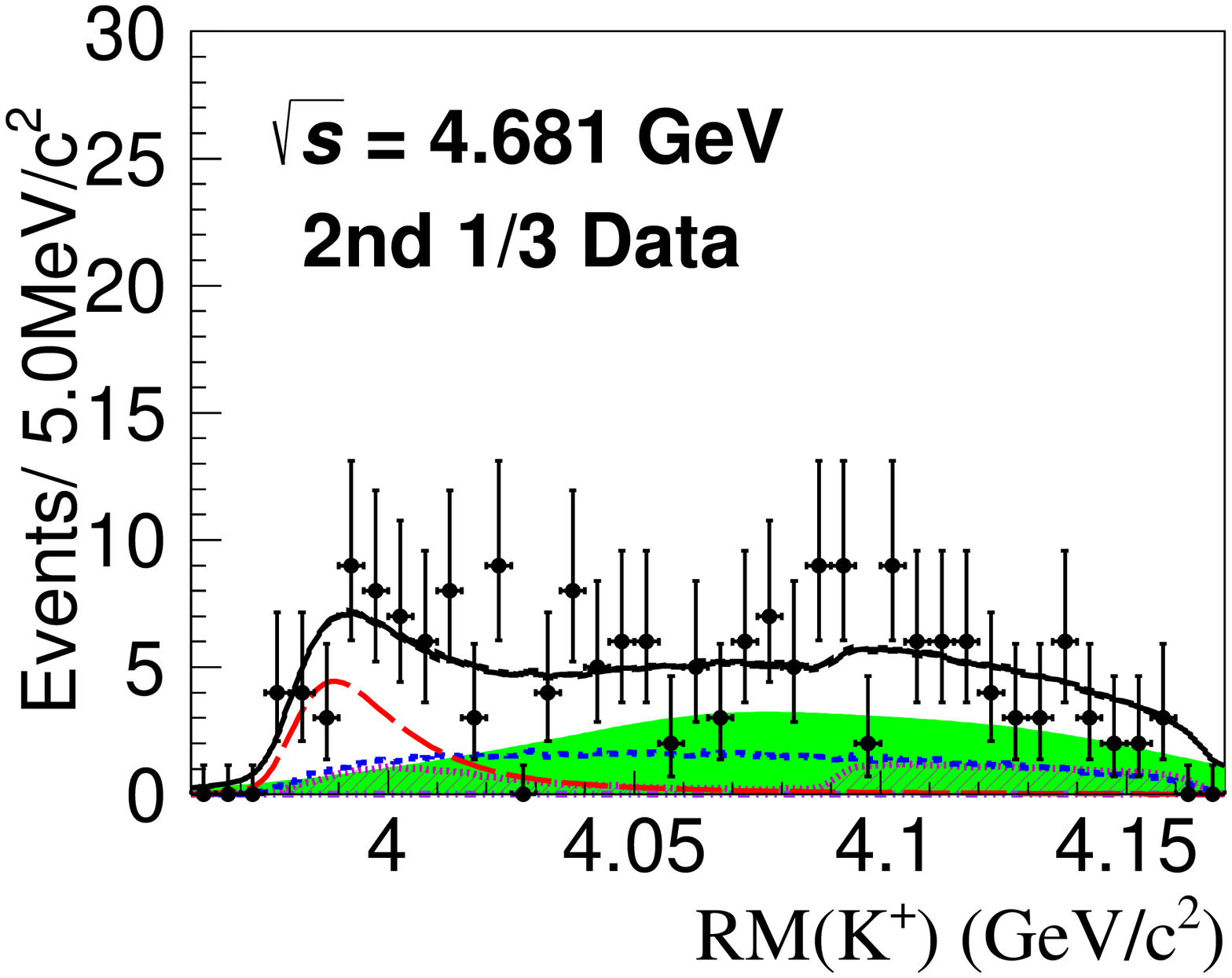}
\includegraphics[width=0.41\linewidth]{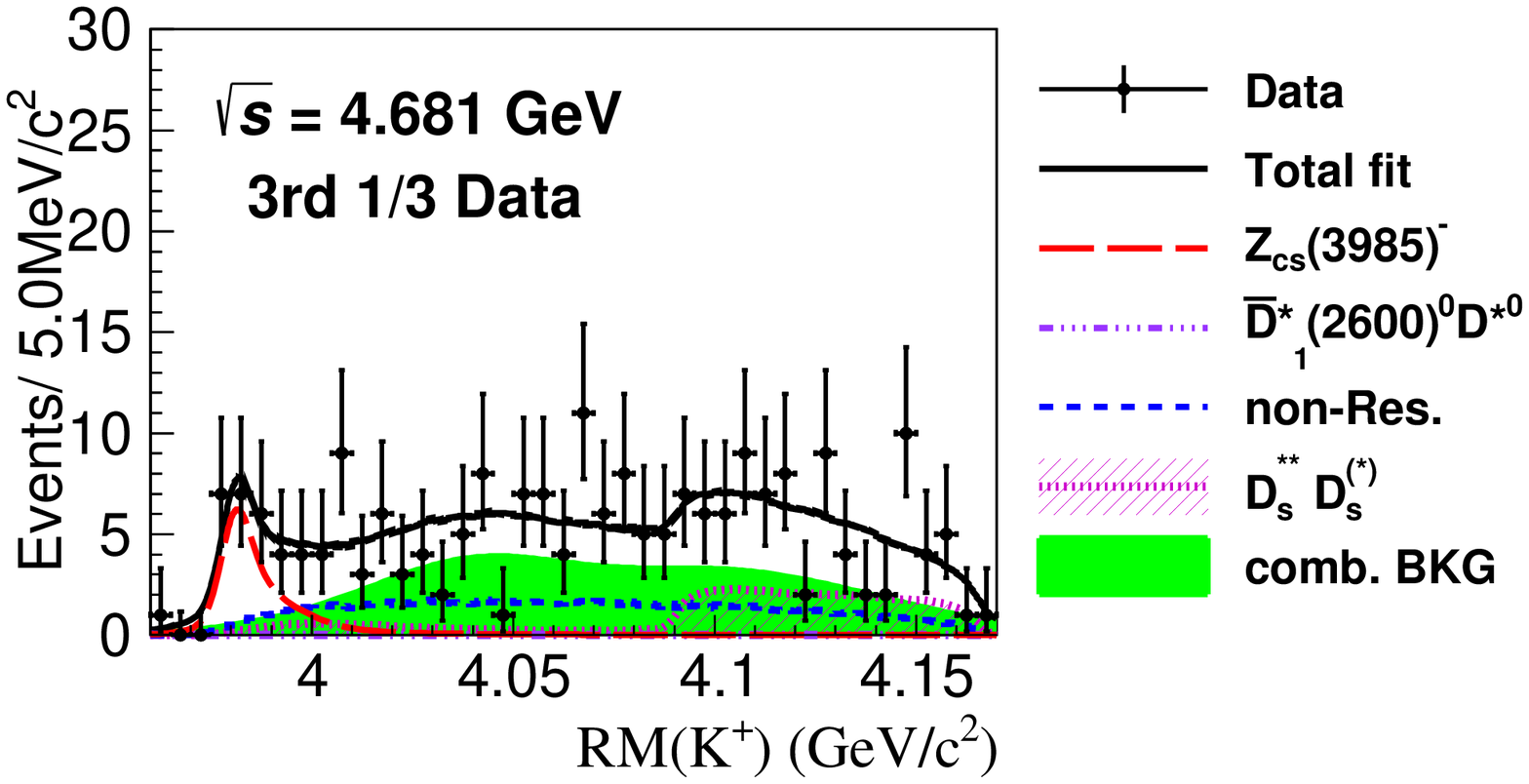}
\vspace{-0.25cm}
\caption{Fit to the $\kaonp$ recoil mass spectra in the first (left),
  second (middle) and third (right) one-third of data set at $\sqrt{s}=4.681 \gev$.}
\vspace{-0.25cm}
\label{fig:4681}
\end{figure}

To avoid potential bias, the analysis strategy is firstly implemented and validated using the first one-third of data set at $\sqrt{s}=\;$4.681~GeV, where the fit result is shown in Fig.~\ref{fig:4681}(left) and given in Table~\ref{tab:4681}. Afterward, we split the two-thirds of data into two parts for consistency check by implementing the same fit procedures, the results of which are depicted in Fig.~\ref{fig:4681}(middle) and (right). The corresponding numerical results are listed in Table~\ref{tab:4681}. The fitted resonance parameters between the 1st and 2nd one-third of data set are consistent within statistical uncertainty, while the comparison between the 1st and 3rd one-third of data set shows that the fitted masses and widths are in agreement within 1.5$\sigma$ and 1.0$\sigma$, respectively. Overall, the three sets of fit results are compatible and we can assume they are due to the same source. Hence, the three parts of data at $\sqrt{s}=\;$4.681~GeV are combined to obtain the nominal fit results listed in Table~\ref{tab:4681}.

\begin{table}[h!]
  \begin{center}
   \caption{Fit results of the $\zcsm$ resonance parameters and cross sections based on the first,  second and third one-third of data set at $\sqrt{s}=4.681 \gev$.}
  \begin{tabular}{lcccc}
      \hline \hline
 Data set      & Mass ($\mevcc$)        & Width ($\mev$)        & $\sigma_{4.681}\cdot\mathcal{B}$(pb)  & Statistical Significance  \\ \hline
 1st one-third & $3987.0^{+2.1}_{-2.4}$ & $6.9^{+6.1}_{-4.1}$   & $5.1^{+1.4}_{-1.2}$ & 4.9$\sigma$  \\
 2nd one-third & $3990.2^{+5.6}_{-5.5}$ & $24.2^{+31.0}_{-12.4}$& $5.0^{+2.3}_{-1.8}$ & 2.9$\sigma$  \\
 3rd one-third & $3980.9^{+2.0}_{-2.2}$ & $4.7^{+9.9}_{-4.7}$   & $2.8^{+1.2}_{-1.0}$ & 3.9$\sigma$  \\ \hline
 nominal       & $3985.2^{+2.1}_{-2.0}$ & $13.8^{+8.1}_{-5.2}$  & $4.4^{+0.9}_{-0.8}$ & 6.3$\sigma$  \\
        \hline\hline
    \end{tabular}
    \label{tab:4681}
  \end{center}
  \end{table}

\clearpage
\section{Calculation of the pole mass and width}
The pole position $m_{\rm pole}(\zcsm)-i\frac{\Gamma_{\rm pole}(\zcsm)}{2}$ is determined by solving the equation
\begin{eqnarray}
 \label{eq:poleEq}
\left\{
\begin{aligned}
& M^{2}-m_{0}^{2}+im_{0}(f\Gamma_{1}(M)+(1-f)\Gamma_{2}(M))=0, \\
& \Gamma_{1}(M)=\Gamma_{0}\cdot\frac{p_{1}}{p_{1}^{*}}\cdot\frac{m_{0}}{M},\\
&
\Gamma_{2}(M)=\Gamma_{0}\cdot\frac{p_{2}}{p_{2}^{*}}\cdot\frac{m_{0}}{M},
\end{aligned}
\right.
\end{eqnarray}
where the input values of $m_0$ and $\Gamma_0$ are taken from the simultaneous fit.
The resonance mass is  above the mass thresholds of the two coupled
channels and the pole position is taken from Riemann sheet III defined
in Ref.~\cite{Badalian:1981xj}. To properly account for their
correlations, a Monte-Carlo method is adopted, in which pseudo data of $m_0$ and $\Gamma_0$ are generated according to the correlation matrix to calculate the pole position.

\end{appendices}


\begin{thebibliography}{99}

  \bibitem{XYZ_review1}
   N.~Brambilla, S.~Eidelman, C.~Hanhart {\it et al.},
  Phys.\ Rep.\ {\bf 873}, 1 (2020).

  \bibitem{XYZ_review2}
  F.~K.~Guo, C.~Hanhart, U.~G.~Mei{\ss}ner, Q.~Wang, Q.~Zhao and B.~S.~Zou,
  Rev.\ Mod.\ Phys.\ {\bf 90}, 015004 (2018).

  \bibitem{XYZ_review3}
  H.~X.~Chen, W.~Chen, X.~Liu and S.~L.~Zhu,
  Phys.\ Rep.\ {\bf 639}, 1 (2016).

  \bibitem{XYZ_review4}
  N.~Brambilla {\em et al.},
  Eur.\ Phys.\ J.\ C {\bf 71}, 1534 (2011).

  \bibitem{XYZ_review5}
  S.~L.~Olsen, T.~Skwarnicki and D.~Zieminska,
  Rev.\ Mod.\ Phys.\ {\bf 90}, 015003 (2018).

  \bibitem{XYZ_review6}
  S.~L.~Olsen,
  Front.\ Phys.\ (Beijing) {\bf 10}, 121-154 (2015).

  \bibitem{Ablikim:2013mio1}
  M.~Ablikim {\it et al.} (BESIII Collaboration),
  Phys.\ Rev.\ Lett.\  {\bf 110}, 252001 (2013).

  \bibitem{Ablikim:2013wzq1}
  M.~Ablikim {\it et al.} (BESIII Collaboration),
  Phys.\ Rev.\ Lett.\  {\bf 111}, 242001 (2013).

  \bibitem{Ablikim:2013xfr1}
  M.~Ablikim {\it et al.} (BESIII Collaboration),
  Phys.\ Rev.\ Lett.\  {\bf 112}, 022001 (2014).

  \bibitem{Ablikim:2013emm1}
  M.~Ablikim {\it et al.} (BESIII Collaboration),
  Phys.\ Rev.\ Lett.\  {\bf 112}, 132001 (2014).

  \bibitem{Ablikim:2014dxl}
  M.~Ablikim {\it et al.} (BESIII Collaboration),
  Phys.\ Rev.\ Lett.\  {\bf 113}, 212002 (2014).

  \bibitem{Ablikim:2015tbp}
  M.~Ablikim {\it et al.} (BESIII Collaboration),
  Phys.\ Rev.\ Lett.\  {\bf 115}, 112003 (2015).

  \bibitem{Ablikim:2015vvn}
  M.~Ablikim {\it et al.} (BESIII Collaboration),
  Phys.\ Rev.\ Lett.\  {\bf 115}, 182002 (2015).

  \bibitem{Ablikim:2015gda}
  M.~Ablikim {\it et al.} (BESIII Collaboration),
  Phys.\ Rev.\ Lett.\  {\bf 115}, 222002 (2015).

  \bibitem{Liu:2013dau}
  Z.~Q.~Liu {\it et al.} (Belle Collaboration),
  Phys.\ Rev.\ Lett.\  {\bf 110}, 252002 (2013);
  Erratum: [Phys.\ Rev.\ Lett.\  {\bf 111}, 019901 (2013)].

  \bibitem{Xiao:2013iha}
  T.~Xiao, S.~Dobbs, A.~Tomaradze and K.~K.~Seth,
  Phys.\ Lett.\ B {\bf 727}, 366 (2013).

  \bibitem{Voloshin:2019ilw}
  M.~B.~Voloshin,
  Phys.\ Lett.\ B {\bf 798}, 135022 (2019).

  \bibitem{Ebert:2008kb}
  D.~Ebert, R.~N.~Faustov and V.~O.~Galkin,
  Eur. Phys. J. C \textbf{58}, 399 (2008).

  \bibitem{Ferretti:2020ewe}
  J.~Ferretti and E.~Santopinto,
  JHEP \textbf{04}, 119 (2020).

  \bibitem{Lee:2008uy}
  S.~H.~Lee, M.~Nielsen and U.~Wiedner,
  J.\ Korean Phys.\ Soc.\  {\bf 55}, 424 (2009).

  \bibitem{Dias:2013qga}
  J.~M.~Dias, X.~Liu and M.~Nielsen,
  Phys.\ Rev.\ D {\bf 88}, 096014 (2013).

  \bibitem{Chen:2013wca}
  D.~Y.~Chen, X.~Liu and T.~Matsuki,
  Phys.\ Rev.\ Lett.\  {\bf 110}, 232001 (2013).

  \bibitem{Ablikim:2009aa}
  M.~Ablikim {\it et al.} (BESIII Collaboration),
  Nucl.\ Instrum.\ Meth.\ A {\bf 614}, 345 (2010).

  \bibitem{Yu:IPAC2016-TUYA01}
   C.~H.~Yu {\it et al.},
  Proceedings of IPAC2016, Busan, Korea, 2016,
  doi:10.18429/JACoW-IPAC2016-TUYA01.

  \bibitem{Ablikim:2019hff}
  M.~Ablikim {\it et al.} (BESIII Collaboration),
  Chin. Phys. C {\bf 44}, 040001 (2020).

  \bibitem{geant4}
  S.~Agostinelli {\it et al.} (GEANT4 Collaboration),
  Nucl.\ Instrum.\ Meth.\ A {\bf 506}, 250 (2003).

  \bibitem{Ablikim:2018qjv}
  M.~Ablikim {\it et al.} (BESIII Collaboration),
  Chin.\ Phys.\ C {\bf 43}, 031001 (2019).

  \bibitem{supple}
  See Supplemental Material for additional analysis information, which includes Ref.~\cite{pdg,Badalian:1981xj}.

  \bibitem{Ablikim:2018jun}
  M.~Ablikim \textit{et al.}  (BESIII Collaboration),
  Phys. Rev. Lett. \textbf{122}, 071802 (2019).

  \bibitem{pdg}
  P.~A.~Zyla \textit{et al.} (Particle Data Group),
  Prog.\ Theor.\ Exp.\ Phys.\ {\bf 2020}, 083C01 (2020).

  \bibitem{Aaij:2019sqk}
  R.~Aaij \textit{et al.} (LHCb Collaboration),
  Phys. Rev. D \textbf{101}, 032005 (2020).

  \bibitem{Aubert:2009ah}
  B.~Aubert \textit{et al.} (BaBar Collaboration),
  Phys. Rev. D \textbf{80}, 092003 (2009).

  \bibitem{Godfrey:2015dva}
  S.~Godfrey and K.~Moats,
  Phys. Rev. D \textbf{93}, 034035 (2016).

  \bibitem{Cranmer:2000du}
  K.~S.~Cranmer,
  Comput.\ Phys.\ Commun.\  {\bf 136}, 198 (2001).

  \bibitem{Gross:2010qma}
  E.~Gross and O.~Vitells,
  Eur. Phys. J. C \textbf{70}, 525 (2010).

  \bibitem{Jegerlehner:2011mw}
  F.~Jegerlehner,
  Nuovo Cim.\ C {\bf 034S1}, 31 (2011).

    \bibitem{Badalian:1981xj}
  A.~M.~Badalian, L.~P.~Kok, M.~I.~Polikarpov and Y.~A.~Simonov,
  Phys. Rept. \textbf{82}, 31 (1982).

\end{thebibliography}
\end{document}